\documentclass[aps,prl,twocolumn,floats,showpacs,superscriptaddress]{revtex4-2}
\usepackage{graphicx,amsmath,amsfonts,amssymb,upgreek,txfonts,color}
\usepackage[colorlinks,linkcolor=blue,citecolor=blue,urlcolor=blue,hyperindex,breaklinks]{hyperref}
\usepackage{color, colortbl}
\definecolor{lightgreen}{rgb}{0.88,1,1}

\begin{document}

\title{Attainable quantum speed limit for N-dimensional quantum systems}
\author{Zi-yi Mai}
\author{Chang-shui Yu}
\email{ycs@dlut.edu.cn}
\affiliation{School of Physics, Dalian University of Technology, Dalian
116024, P.R. China}
\date{\today }

\begin{abstract}
Quantum speed limit (QSL) is a fundamental concept in quantum mechanics and provides a lower bound on the evolution time. The attainability of QSL, greatly depending on the understanding of QSL, is a long-standing open problem especially for high-dimensional systems.  In this paper, we solve this problem by establishing a QSL suitable and attainable for both open and closed quantum systems based on a new proposed state distance. It is shown that given any initial state in a certain dimension, our QSL bound can always be saturated by unitary and non-unitary dynamics, and for any given Hamiltonian for a unitary evolution, a pair of states always exists, saturating the bound.  As applications, we demonstrate the QSL time attained by various physical settings. This paper will shed new light on the QSL problems. \end{abstract}
\maketitle

\section{Introduction}

Quantum speed limit (QSL) is a fundamental concept in quantum mechanics. It
provides a lower bound on the evolution time between two states \cite{frey2016quantum, Deffner_2017}.  QSL has been attracting increasing interest and indicates potential consequences in various scenarios such as quantum metrology \cite%
{giovannetti2011advances,PhysRevLett.96.010401,PhysRevLett.109.233601},
quantum optimal control \cite%
{PhysRevLett.103.240501,PhysRevLett.117.030802,PhysRevA.84.022305,Poggi_2013,PhysRevA.92.062110}%
, quantum measurement \cite{Garcia_Pintos_2019}, thermometry \cite%
{Campbell_2018} and so on.

The concept of QSL was initially introduced by Mandelstam and Tamm \cite%
{Mandelstam1991} in 1945 for the unitary evolution between two orthogonal pure states. They found a lower bound (MT) $\tau_{MT}^\bot=\pi\hbar/(2\Delta E)$ on evolution time with $\Delta E$ denoting the energy fluctuation of the closed system.
Margolus and Levitin proposed another bound (ML)
  based on the average energy as $%
\tau_{ML}^\bot=\pi\hbar/2 (E-E_0)$ \cite%
{MARGOLUS1998188} with $E$ denoting the average energy and $E_0$ representing system's ground-state energy.  Later, QSL was generalized in many different  aspects, including
time-dependent Hamiltonian \cite{PhysRevLett.65.1697},  incompletely
distinguishable state pairs \cite%
{fleming1973unitarity,uffink1993rate,Deffner_2013,frey2016quantum}, mixed states \cite%
{PhysRevLett.120.060409,Huang_2022,PhysRevA.67.052109,PhysRevA.98.042132},
non-unitary dynamics \cite%
{PhysRevLett.110.050402,Campaioli2019tightrobust,PhysRevA.95.022115,PhysRevLett.110.050403}%
,  and QSLs in the geometric perspectives \cite%
{PhysRevA.103.022210,MONDAL20161395,PhysRevLett.123.180403,PhysRevLett.72.3439,PhysRevLett.65.1697,PhysRevX.6.021031,PhysRevA.82.022107}.
QSL has been widely studied in open systems to explore the speed-up mechanism induced by the systems' features  such as the non-Markovianity\cite%
{PhysRevLett.111.010402,sun2015quantum,PhysRevA.91.022102}, entanglement
\cite{PhysRevA.91.022102,PhysRevA.78.042305,Zander_2007}, coherence \cite%
{PhysRevA.102.053716,zhang2014quantum,wu2020quantum,PhysRevA.93.052331,MONDAL2016689}
and so on \cite{PhysRevA.103.062204,Hou_2015,PhysRevA.89.012307,Wu_2015,PhysRevA.95.052104}.  In recent years,  QSL has been investigated in more general cases like quantum resource variation  \cite{Campaioli_2022}, changing the expected value of observables \cite{PhysRevX.12.011038, PhysRevA.106.042436}, and even in classical systems  \cite%
{PhysRevLett.120.070402,PhysRevLett.120.070401}.

QSL is a type of time-optimal problem with some constraints.  Even though various significant bounds for QSL time, as mentioned above, have gained a deep understanding of the evolution of quantum systems,   their attainability is usually unachievable or challenging to verify except for some attainable ones in two-level systems. Constructing the attainable QSL bound seems complicated, as it is not only directly related to the potential constraints but also depends on different understandings of the QSL and attainability.  Both MT and ML bounds typically indicate
the minimal evolution time for a given `energy scale' and are simultaneously saturated for two-level systems. Maximizing different bounds, for example, the MT and ML bounds gives tight bounds that reveal what quantities determine the minimal evolution time between a pair of quantum states. Recently, the attainable or the tight bounds on the QSL times have been proposed from many perspectives for closed or open systems \cite{Campaioli2019tightrobust, PhysRevA.108.052207}.  However, it is still an open problem to establish a QSL that is attainable for any given $N$-dimensional quantum systems and suitable for both closed and open systems.

In this paper, we provide an alternative solution to the above problem by presenting an attainable QSL based on our defined state distances. It is shown that our QSL bound
can be applied and attained in both closed and open systems.  For any $N$-dimensional quantum state, one can always find an unitary dynamics and an non-unitary dynamics driving the system to evolve along their corresponding geodesic,  respectively.  Given a Hamiltonian for an unitary evolution, one can also find a pair of states saturating the QSL bound. Besides the rigorous proofs, we have also applied our results to concrete physical systems and demonstrated the evolution time with the optimal dynamics.

\section{The state distance}
First, let's introduce a key
distance of quantum states. For any quantum state $\rho$, one can always
establish an injective mapping $F_\alpha(\rho)$ as
\begin{equation}
F_\alpha(\rho)=\rho-\frac{1+\alpha-\mathrm{Tr}\rho^2}{N}\mathbb{I},\label{map}
\end{equation}
where $\alpha\in (1/N,1]$ is a constant and $\mathbb{I}$ is the identity of
the same dimension of the density $\rho$.  One can find from Supplemental Material that the constant $\alpha$
garantees the injective mapping $\rho\rightarrow F_\alpha (\rho)$. Thus, we
can define the distance between the two states as follows:

The distance of two states $\rho$
and $\sigma$
\begin{equation}  \label{dis1}
D_\alpha(\rho,\sigma)=\arccos\left\langle \bar{F}_\alpha(\rho),\bar{F}%
_\alpha(\sigma)\right\rangle,
\end{equation}
where $\bar{F}_\alpha=F_\alpha/\left\vert F_\alpha\right\vert$, $%
\left\langle A,B\right\rangle=\mathrm{Tr}\{A^\dagger B\}$ denotes the Hilbert-Schmidt (H-S) inner product for any operators $%
A$, $B$ and $\left\vert A\right\vert=\sqrt{\left\langle
A, A\right\rangle}$.

It can be found that $D_\alpha$ in Eq. (\ref{dis1}) satisfies all the
properties for a good distance. $D_\alpha (\rho,\sigma)$ is symmetric under
exchanging $\rho$ and $\sigma$. It is non-negative, and especially, $%
D_\alpha (\rho,\sigma)=0$ if and only if $\rho=\sigma$ which is due to the
injective mapping $\rho\rightarrow F_\alpha (\rho)$. A simple algebra can
show that the distance is invariant under a unitary transformation $U$
simultaneously performed on both matrices, i.e., $D_\alpha(\rho,\sigma)=D_%
\alpha(U\rho U^\dagger,U\sigma U^\dagger)$. The most important is that $%
D_\alpha (\rho,\sigma)$ satisfies the triangular inequality due to the injection given in Eq. (\ref{map}) (a detailed proof is given
in Supplemental Material).

The above defined distance $D_{\alpha }$ is obviously suitable for any pair
of states; however, we have to construct another one for our particular
purpose in the paper. For any $N$-dimensional density matrix $\rho $, it
can always be written in the eigendecomposition, i.e., $\rho
=\Phi \Lambda \Phi ^{\dagger }$, where $\Phi $ is an unitary matrix with its
columns corresponding the eigenvector, and $\Lambda $ is a diagonal matrix
with its diagonal entries corresponding to the eigenvalues.  With the action
of the discrete Fourier transformation (DFT) $U_{mn}^{F}=\frac{1}{\sqrt{N}}%
e^{\frac{2mn\pi i}{N}}$, one can find that any diagonal matrix $\Lambda $
can be converted to the matrix $U^{F}\Lambda \left( U^{F}\right) ^{\dagger }$
with the uniform diagonal entries $1/N$. Let's introduce a projector as
\begin{equation}
P_{ij}\left( \Phi \right) =\Phi P_{ij}\Phi ^{\dag },P_{ij}=\left\vert
i_{F}\right\rangle \left\langle i_{F}\right\vert +\left\vert
j_{F}\right\rangle \left\langle j_{F}\right\vert ,i\neq j,
\end{equation}%
where $\left\vert i_{F}\right\rangle $ denotes $i$th column of the matrix
$U^{F}$, and we use the subscript and superscript 'F' to represent DFT
throughout the paper. Thus, based on the projectors, we can establish a new
matrix (named \textit{Projective Matrix} if not make confusions) as
\begin{equation}
\lbrack \rho ]_{ij}=P_{ij}\left( \Phi \right) \rho P_{ij}\left( \Phi \right)
+\frac{1}{N}(\mathbb{I}_{N}-P_{ij}\left( \Phi \right) ),  \label{pro}
\end{equation}%
where $\mathbb{I}_{N}$ is the $N$-dimensional identity and $[\cdot ]_{ij}$
denotes the particular projection operation. It is obvious that given a
density matrix $\rho $, one can construct $\frac{N(N-1)}{2}$ projective
matrices $[\rho ]_{ij}$ with respect to the different choices of $(i,j)\in
\lbrack 1, N]$. In addition, the projective matrix keeps the properties of a
density matrix, and additionally owns a good property that
\begin{equation}
\left[ U\rho U^{\dag }\right] _{ij}=U[\rho ]_{ij}U^{\dag }  \label{pro2}
\end{equation}%
holds for any unitary operation $U$. With these established projective
matrices, one can define the distance in the following way:

For $N$-dimensional density matrices
$\rho$ and $\sigma$ with $\Phi$ and $\Psi$
denoting the eigenvector unitary matrices,
their corresponding projective matrix sets are denoted by $%
\left\{[\rho]_{ij}\right\}$ and $\left\{[\sigma]_{ij}\right\}$,
respectively. The distance of $\rho$ and $\sigma$
in the frame of $\{\Phi,\Psi\}$ is defined by
\begin{equation}
\bar{D}_{\alpha}^{\{\Phi,\Psi\}}(\rho,\sigma)=\sum_{i\neq
j,i,j=1}^ND_{\alpha_{ij}}([\rho]_{ij},[\sigma]_{ij}),  \label{dis2}
\end{equation}
where $\alpha$ different from that in Eq. (\ref{dis1})
represents a constant matrix for convenience with $\alpha_{ij}\in (1/N,1]$
denoting the $(i,j)$ matrix element and its diagonal
entires undefined.  $\alpha_{ij}$ can be freely chosen for some
particular purposes.

It is not difficult to see that Eq. (\ref{dis2}) generalizes
Eq. (\ref{dis1}). The two measured high-dimensional quantum states are cut and
mended into a series of state pairs in a particular way. $D_\alpha$ measures
the distance of every pair. $\bar{D}_\alpha^{\{\Phi,\Psi\}}$ collects the
contributions of $D_\alpha$ for all the pairs. In this sense, one can find
that $\bar{D}_\alpha^{\{\Phi,\Psi\}}$ is equivalent to $D_\alpha$ for a
qubit state. Next, one can see that Eq. (\ref{dis2}) also defines a valid
state distance. At first, one can easily find that the distance is symmetric
in the sense of $\bar{D}_\alpha^{\{\Phi,\Psi\}}(\rho,\sigma)=\bar{D}%
_\alpha^{\{\Phi,\Psi\}}(\sigma,\rho)$. Since $D_\alpha\geq 0$, it is natural
that $\bar{D}_\alpha^{\{\Phi,\Psi\}}\geq 0$. In particular, $\bar{D}%
_\alpha^{\{\Phi,\Psi\}}= 0$ means $\rho=\sigma$, which can be shown as
follows. $\bar{D}_\alpha^{\{\Phi,\Psi\}}= 0$ is equivalent to $%
D_{\alpha_{ij}}([\rho]_{ij})=D_{\alpha_{ij}}([\sigma]_{ij})$ for all
potential $(i,j)$. Namely, $[\rho]_{ij}=[\sigma]_{ij}$ holds for all the
potential $(i,j)$ due to the properties of $D_\alpha$, which implies $%
\sum_{ij}[\rho]_{ij}=\sum_{ij}[\sigma]_{ij}$. According to Eq. (\ref{pro}),
one can find that $\sum_{ij}[X]_{ij}=2X+(N-1-\frac{2}{N})\mathbb{I}_N$ with $%
X=\rho,\sigma$, which immediately yields $\rho=\sigma$. Additionally, based
on Eq. (\ref{pro2}), $\bar{D}_\alpha^{\{\Phi,\Psi\}}(\rho,\sigma)$ inherits
the invariance under an unitary operation $U$ simultaneously performed on $%
\rho$ and $\sigma$. The property of the triangular inequality of
$D_\alpha$ is also inherited naturally.

Note that the $U^F$ plays a key role in defining the state distance.  However,  the matrix $U^F$
is not the unique choice to achieve a valid definition.  For example,  one can attach the
 permutation matrix $U^P$ to replace $U^F$ as $U^FU^P\rightarrow U^F$ to get another distance.  Such a nonuniqueness could lead to nonunique results, especially for orthogonal states, which can be easily solved by maximizing over all the permutation matrices as \begin{equation}\tilde{D}_\alpha  ^{\left\{\Phi,\Psi\right\}}\left(\rho,\sigma\right)=\max_{U^P}\bar{D}_\alpha ^{\left\{\Phi,\Psi\right\}}\left(\rho,\sigma\right).\end{equation}
It can be found that $\tilde{D}_\alpha  ^{\left\{\Phi,\Psi\right\}}\left(\rho,\sigma\right)$ is also a good state distance.

\section{Quantum speed limit}
Up to now, we have successfully constructed
several distances. With these distances, one can go forward to study the
quantum speed limit. Suppose the dynamics of a quantum system is written in the general form
\begin{equation}
\dot{\rho}_t=\mathcal{L}(\rho_t),  \label{dynamics}
\end{equation}
where $\rho_t$ means the quantum state at the moment $t$ and $\mathcal{L}%
(\cdot)$ represents general dynamics, including unitary and non-unitary
evolutions. Let $\rho_{t+dt}$ be the state after the evolution of
infinitesimal duration $dt$ from $\rho_t$.

Firstly, we employ $D_\alpha$ given in Eq. (\ref{dis1}) to measure the distance of the states $\rho_t$
and $\rho_{t+dt}$, which can be written as
\begin{eqnarray}
&&D_\alpha^2(\rho_t,\rho_{t+dt})=2-2\cos D_\alpha(\rho_t,\rho_{t+dt})  \notag
\\
&&=2-2\left\langle \bar{F}_\alpha(\rho_t),\bar{F}_\alpha(\rho_{t+dt})\right%
\rangle  \notag \\
&&=\left\langle \bar{F}_\alpha(\rho_{t+dt})-\bar{F}_\alpha(\rho_{t}),\bar{F}%
_\alpha(\rho_{t+dt})-\bar{F}_\alpha(\rho_{t})\right\rangle  \notag \\
&&=\left\langle d\bar{F}_\alpha(\rho_{t}),d\bar{F}_\alpha(\rho_{t})\right%
\rangle=\left\vert d\bar{F}_\alpha(\rho_{t})\right\vert^2,  \label{metric}
\end{eqnarray}
where `$d$' represents the differential of the following function. Thus one
can directly arrive at our first QSL time for a dynamics as Eq. (\ref%
{dynamics}) as follows.

\textbf{Theorem 1}.-The needed time $\tau$ for the evolution from $%
\rho_0$ to $\rho_\tau$ governed by the dynamics Eq. (\ref{dynamics}) is
lower bounded by
\begin{equation}  \label{QSL1}
\tau\geq\tau_\alpha=\frac{D_\alpha\left(\rho_0,\rho_\tau\right)}{%
\left\langle\left\vert\frac{d\bar{F}_\alpha}{dt}\right\vert\right\rangle_\tau%
},
\end{equation}
where the average $\left<\cdot\right\rangle_\tau=\frac{1}{\tau}%
\int^\tau_0(\cdot) dt$ and
\begin{equation}  \label{mc}
\begin{split}
\left\vert\frac{d\bar{F}_\alpha}{dt}\right\vert^2=&\frac{\mathrm{Tr}%
\dot\rho_t^2}{\left\vert F_\alpha(\rho_t)\right\vert^2}- \frac{1-\frac{4}{N}%
\left(\alpha-\frac{1}{N}\right)}{\left\vert F_\alpha(\rho_t)\right\vert^4}%
\left(\mathrm{Tr}\rho_t\dot\rho_t\right)^2.
\end{split}%
\end{equation}

 The proof of Eq. (\ref{QSL1}) is quite straightforward. Substituting the metric
given in Eq. (\ref{metric}) into the triangular inequality of the distance $%
D_\alpha$, one can directly obtain
\begin{equation}
D_\alpha(\rho_0,\rho_\tau)\leq \int_0^\tau D_\alpha(\rho_t,\rho_{t+dt})
=\int_0^\tau \left\vert \frac{d}{dt}\bar F_\alpha(\rho_t)\right\vert dt,
\end{equation}
which is exactly Eq. (\ref{QSL1}), and  Eq. (\ref{mc}) can be directly obtained
by calculating the derivative of $\bar{F}_\alpha$ over $t$.

Eq. (\ref{QSL1}) gives a QSL based on the distance $D_\alpha\left(\rho_0,\rho_\tau\right)$. However,
the attainability is limited, which won't be discussed here. What we want to emphasize is that $\bar{D}^{\Phi,\Psi}_\alpha$ coming from $D_\alpha\left(\rho_0,\rho_\tau\right)$ can produce a better QSL, which is our main result.

Secondly, we use $\bar{D}^{\Phi,\Psi}_\alpha$ in Eq. (\ref{dis2}) to measure the distance of $%
\rho_t$ and $\rho_{t+dt}$. For every $(i,j)$, following Eq. (\ref{metric})
one can easily obtain that
\begin{equation}
\bar{D}_{\alpha_{ij}}\left(\left[\rho_t\right]_{ij},\left[\rho_{t+dt}\right]%
_{ij}\right)=\left\vert d \bar{F}^{\Phi_t}_{\alpha_{ij}}\left(\left[\rho_t%
\right]_{ij}\right)\right \vert,
\end{equation}
where $\Phi_t$ is the eigenvector unitary matrix of $\rho_t$. Thus the
distance of $\rho_t$ and $\rho_{t+dt}$ reads
\begin{equation}
\bar{D}^{\Phi_t,\Phi_{t+dt}}_\alpha(\rho_t,\rho_{t+dt})=\sum_{i,j}\left\vert
d \bar{F}^{\Phi_t}_{\alpha_{ij}}\left(\left[\rho_t\right]_{ij}\right)\right
\vert.  \label{met2}
\end{equation}
Here, we would like to emphasize that a particular continuous evolution
trajectory of $\Phi_t$ is chosen in Eq. (\ref{met2}) if the degeneracy is
present. Analogous to the QSL in Eq. (\ref{QSL1}), we can give a second QSL as follows.

\textbf{Theorem 2}.-The needed time $\tau$ for the evolution from $%
\rho_0$ to $\rho_\tau$ governed by the dynamics Eq. (\ref{dynamics}) is
lower bounded by
\begin{equation}  \label{QSL2}
\tau\geq\tau_{\mathrm{QSL}}=\frac{\bar{D}_\alpha^{\Phi_0,\Phi_\tau}\left(\rho_0,\rho_%
\tau\right)}{\sum_{i\neq j}\left\langle\left\vert\frac{d\bar{F}%
^{\Phi_t}_{\alpha_{ij}}\left(\left[\rho\right]_{ij}\right)}{dt}%
\right\vert\right\rangle_\tau}
\end{equation}
where $\Phi_\tau=\Phi_0+\int_0^\tau {d\Phi_t}$, and $\alpha_{ij}=\max\left\{\mathrm{Tr}\left[\rho_0\right]_{ij}^2,\mathrm{Tr}\left[\rho_\tau\right]_{ij}^2\right\}$ can be taken for simplicity.

The proof of Eq. (\ref{QSL2}) is analogous to the proof of Eq. (\ref{QSL1}), so it is omitted
here. The matrix element $\alpha_{ij}$ can be arbitrarily chosen, as mentioned in Eq. (\ref{dis2}). However, we'd like to suggest the maximal purities of the initial and final states, which will lead to a simple expression and a good tightness.

Since Eq. (\ref{dis2}) is a generalization of Eq. (\ref{dis1}), the QSL in Eq. (\ref{QSL2}) is
naturally a generalization of Eq. (\ref{QSL1}). They are completely consistent
with each other for a qubit system. In addition, one can also employ $\tilde{D}_\alpha^{\left\{\Phi,\Psi\right\}}$ to measure the state distance. The corresponding QSL can be naturally obtained by replacing $\bar{D}$ in Eq. (\ref{QSL2}) by $\tilde{D}$, meanwhile $[\rho]_{ij}$ in the denominator should be updated by the permutation operation $U^F$ optimal for $\tilde{D}$, which is applicable for all the following results.  We won't repeat it.

\section{Applications and attainability for closed systems}

As applications, we first employ the QSL in Eq. (\ref{QSL2}) to closed systems under the
unitary evolution. Suppose a state $\rho_0$ undergoes a unitary dynamics as
\begin{equation}  \label{U_path}
\rho_t=U_t \rho_0 U_t^\dagger
\end{equation}
with $U_t=e^{-i\int_0^tdt^{\prime }H_{t^{\prime }}}$ and $H_t$ the
Hamiltonian of the system. According to Eq. (\ref{QSL2}), one can immediately
obtain the QSL time bounded in the following form.

\textbf{Corollary 3}.-The needed time $\tau $ for an $N$%
-dimensional closed system evolving from $\rho _{0}$ \ to $%
\rho _{\tau }$ driven by the Hamiltonian $H_{t}$ is lower
bounded by
\begin{equation}
\tau \geq \tau _{\mathrm{QSL}}=\frac{\bar{D}_{\alpha }^{\Phi ,U_{t}\Phi }\left(
\rho _{0},\rho _{\tau }\right) }{\sum_{i\neq j}\frac{\sqrt{2}}{\sqrt{\mathrm{%
Tr}\left[ \rho _{0}\right] _{ij}^{2}-\frac{1}{N}}}\left\langle \Delta
E\left( \left[ \rho _{t}\right] _{ij}\right) \right\rangle _{\tau }},
\label{U_QSL}
\end{equation}%
where $\Phi $ is the eigenvector unitary matrix for $\rho
_{0}$, $U_{t}$ is defined in Eq. (\ref{U_path}), $\alpha _{ij}=%
\mathrm{Tr}\left[ \rho _{0}\right] _{ij}^{2}$ and
\begin{equation}
\Delta E\left( \left[ \rho _{t}\right] _{ij}\right) =\sqrt{\mathrm{Tr}%
H_{t}^{2}\left[ \rho _{t}\right] _{ij}^{2}-\mathrm{Tr}\left( H_{t}\left[
\rho _{t}\right] _{ij}\right) ^{2}}.  \label{vari}
\end{equation}

The proof of Eq. (\ref{U_QSL}) can be shown as follow: Based on the establishment of the projective matrices, one
can find that the projective matrices for $\rho_t$ and $\rho_0$ have the
relation as $[\rho_t]_{ij}=U_t[\rho_0]_{ij} U_t^\dag$, so the purity of the
projective matrices doesn't change, i.e., $\mathrm{Tr}\left[\rho_t\right]%
_{ij}^2=\mathrm{Tr}\left[\rho_0\right]_{ij}^2$, and
$
\frac{d}{dt}\left[\rho_t\right]_{ij}=-i\left[H_t,\left[\rho_t\right]_{ij}%
\right].
$
Substituting $\left[\rho_t\right]_{ij}$ into Eq. (\ref{pro}), one can obtain
$\left\vert F^{\Phi_t}_{\alpha_{ij}}\left(\left[\rho_t\right]_{ij}\right)\right\vert=%
\sqrt{\mathrm{Tr}\left[\rho_0\right]_{ij}^2-1/N}$ with $\alpha_{ij}=\mathrm{%
Tr}\left[\rho_0\right]_{ij}^2$. Thus, the corresponding metric for the
projective matrices reads
\begin{equation}  \label{U_metric}
\left\vert \frac{d}{dt}\bar F^{\Phi_t}_{\alpha_{ij}}\left(\left[\rho_t\right]%
_{ij}\right)\right\vert^2=2\frac{\Delta E^2\left(\left[\rho_t%
\right]_{ij}\right)}{\mathrm{Tr}\left[\rho_0\right]_{ij}^2-1/N},
\end{equation}
where $\Delta E\left(\left[\rho_t\right]_{ij}\right)=\sqrt{\mathrm{Tr}H_t^2%
\left[\rho_t\right]^2_{ij}-\mathrm{Tr}\left(H_t\left[\rho_t\right]%
_{ij}\right)^2}$. Substituting Eq. (\ref{U_metric}) into Eq. (\ref{QSL2}),
one can directly arrive at the QSL.

One can note that if $H_t$ is time-independent, we have $\Delta E\left(\left[%
\rho_t\right]_{ij}\right)=\Delta E\left(\left[\rho_0\right]_{ij}\right)$. If $\rho_0$ is a pure state of qubits, $\Delta E$ will be the
energy fluctuation. Next, we will demonstrate an application of QSL in Eq. (\ref{U_QSL})
in a concrete system, one will find that the QSL can be attainable
through unitary evolution.

Let
\begin{equation}\label{opt_rho0}
\rho _{0}=\Phi \Lambda _{0}\Phi ^{\dag }
\end{equation}
be the initial state with the eigendecomposition. If the system takes the Hamiltonian
\begin{equation}\label{opt_H}
H=\sum_{i=1}^{N}E_{i}\Phi \left\vert i_{F}\right\rangle \left\langle
i_{F}\right\vert \Phi ^{\dag },
\end{equation}
the QSL in Eq. (\ref{U_QSL}) is
attainable. Namely, the system will evolve along the geodesic.   On the contrary, given a Hamiltonian $
\tilde{H}=\Psi \Lambda_E\Psi^\dag$ in the eigendecomposition form,  one can also find a pair of states
saturating the QSL bound.

It can be verified as follow: Considering the evolution operator $U_t=e^{-iHt}=\sum_{i=1}^{N}e^{-iE_{i}t}\Phi \left\vert i_{F}\right\rangle \left\langle
i_{F}\right\vert \Phi ^{\dag }$,  it is natural $[U_t,H]=0$ and $\Delta E\left(\left[\rho_t\right]_{ij}\right)=\Delta E\left(\left[\rho_0\right]_{ij}\right)$.  Note that $\left[\rho _{0}\right]_{ij}$ can be rewritten as
\begin{eqnarray}&&\left[\rho _{0}\right]_{ij}=P_{ij}\left( \Phi \right) \rho
_{0}P_{ij}\left( \Phi \right)  +\frac{1}{%
N}(\mathbb{I}_{N}-P_{ij}\left( \Phi \right) )\notag\\
&&=\Phi\left(\lambda_{ij}\left\vert i_F\right\rangle\left\langle j_F\right\vert+\lambda_{ij}^*\left\vert j_F\right\rangle\left\langle i_F\right\vert+\frac{\mathbb{I}_N}{N}\right)\Phi^\dag,\label{rrr}
\end{eqnarray}
where $\lambda _{ij}$ is the $(ij)$th element of the matrix $U^{F}\Lambda
_{0}\left( U^{F}\right) ^{\dagger }$, then
$\left[ \rho _{0}\right] _{ij}^{2} $ can be given as
\begin{eqnarray*}
\left[ \rho _{0}\right] _{ij}^{2}
&=& \left\vert \lambda_{ij}\right\vert^2 P_{ij}\left(\Phi\right)
+\frac{\mathbb{I}
_{N}}{N^2} \notag\\
&+&\frac{2}{N}\Phi\left(\lambda _{ij}\left\vert i_{F}\right\rangle
\left\langle j_{F}\right\vert
+\lambda _{ij}^{\ast }\left\vert
j_{F}\right\rangle \left\langle i_{F}\right\vert \right)\Phi ^{\dagger } ,\end{eqnarray*}%
Obviously,
$
\mathrm{Tr}\left[\rho_0\right]_{ij}^2-\frac{1}{N}=2\left\vert\lambda_{ij}\right\vert^2.
$
Considering $H^{2}=
\sum_{k=1}^{N}E_{k}^{2}\Phi\left\vert k_{F}\right\rangle \left\langle
k_{F}\right\vert \Phi ^{\dag }$, one can arrive at
\begin{eqnarray}
&&\mathrm{Tr}H^{2}\left[ \rho _{0}\right] _{ij}^{2}=\left\vert \lambda_{ij}\right\vert^2\left(E_i^2+E_j^2\right)+\frac{1}{N^2}\underset{k}{\sum}E_k^2,\\
 &&\mathrm{Tr}\left( H\left[ \rho _{0}\right] _{ij}\right) ^{2}=\mathrm{Tr}%
\left( E_i\lambda_{ij} \left\vert i_{F}\right\rangle
\left\langle j_{F}\right\vert
+E_j\lambda _{ij}^{\ast }\left\vert
j_{F}\right\rangle \left\langle i_{F}\right\vert   \right.\notag\\
&&+\left.\sum_{k=1}^{N}\frac{E_{k}}{N}\left\vert k_{F}\right\rangle \left\langle k_{F}\right\vert  \right) ^{2}
=2E_{i}E_{j}\left\vert \lambda
_{ij}\right\vert ^{2}+\frac{1}{N^{2}}\sum_k E_k^2,
\label{U_variance2_attain}
\end{eqnarray}%
which yields
$
\Delta E\left(\left[\rho_t\right]_{ij}\right)=\left\vert E_{i}-E_{j}\right\vert \left\vert \lambda
_{ij}\right\vert .  $
Thus, one can calculate the denominator of Eq. (\ref{U_QSL}) is $\sum_{ij}\left\vert E_i-E_j\right\vert$.

In addition, based on the expression of $U_t$ and Eq. (\ref{rrr}),  one can see that
\begin{eqnarray}
&&\mathrm{Tr}\left(\left[\rho_0\right]_{ij}\left[\rho_\tau\right]_{ij}\right)=\mathrm{Tr}\left(\left[\rho_0\right]_{ij}U_\tau\left[\rho_0\right]_{ij}U_\tau^\dag\right)\notag\\
&&=\mathrm{Tr}\left[\rho_0\right]_{ij}\left(e^{-i(E_i-E_j)\tau}\lambda_{ij}\left\vert i_F\right\rangle\left\langle j_F\right\vert+h.c.\right)+\frac{1}{N}\notag\\
&&=\left\vert\lambda_{ij}\right\vert^2\left(e^{-i(E_i-E_j)\tau}+e^{i(E_i-E_j)\tau}\right)+\frac{1}{N}.
\end{eqnarray}
Hence the distance of $\rho_0$ and $\rho_\tau$ reads
\begin{eqnarray}  \label{U_attain}
&&\bar{D}_\alpha^{\Phi,U_t\Phi}(\rho_0,\rho_\tau)=\sum_{i\neq j}D_{\alpha_{ij}}\left(\left[\rho_0\right]_{ij},\left[\rho_\tau\right]_{ij}\right)\notag \\
&&=\sum_{i\neq j}\arccos\frac{\mathrm{Tr}\left[\rho_0\right]_{ij}\left[%
\rho_\tau\right]_{ij}-\frac{1}{N}}{\mathrm{Tr}\left[\rho_0\right]^2_{ij}-%
\frac{1}{N}}\notag \\
&&=\sum_{i\neq j}\arccos\frac{\left\vert \lambda_{ij}\right\vert^2\left[%
e^{-i(E_i-E_j)\tau}+e^{i(E_i-E_j)\tau}\right]}{2\left\vert\lambda_{ij}\right%
\vert^2} \notag\\
&&=\sum_{i\neq j}\left\vert E_i-E_j\right\vert\tau,
\end{eqnarray}
combined with the above denominator value, the QSL time is $\tau_{QSL}=\tau$. This means that the given dynamics saturates the QSL inequality (\ref{U_QSL}).

On the contrary, if $
\tilde{H}=\Psi \Lambda_E\Psi^\dag$  is given, one can immediately find that the intial state $\tilde{\rho}_0=\Phi U^F\Psi^\dag\Lambda_0\Psi \left(U^F\right)^\dag\Phi^\dag$ can saturate the QSL bound, which can be shown by substitution into the above proof.  The proof is completed.

\begin{figure}
\includegraphics[width=0.46\linewidth]{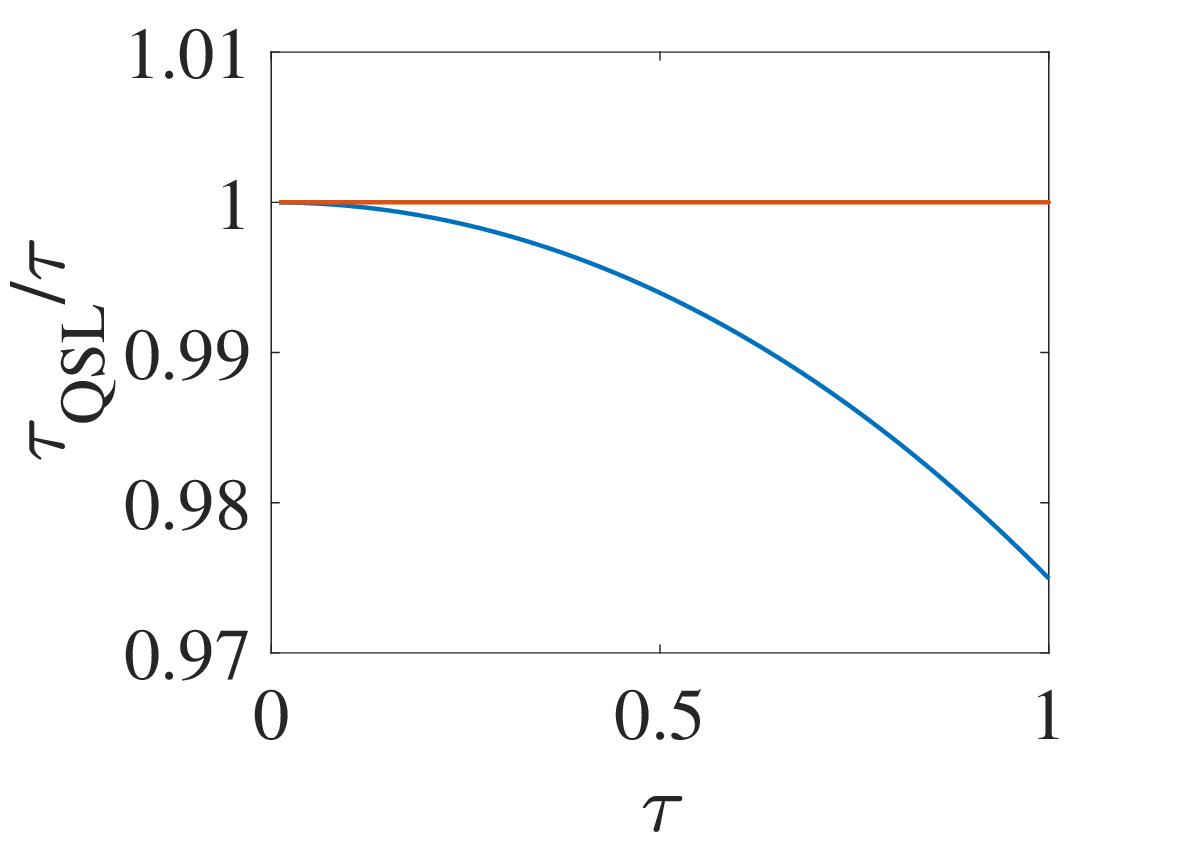}
\label{jcv} \includegraphics[width=0.46%
\linewidth]{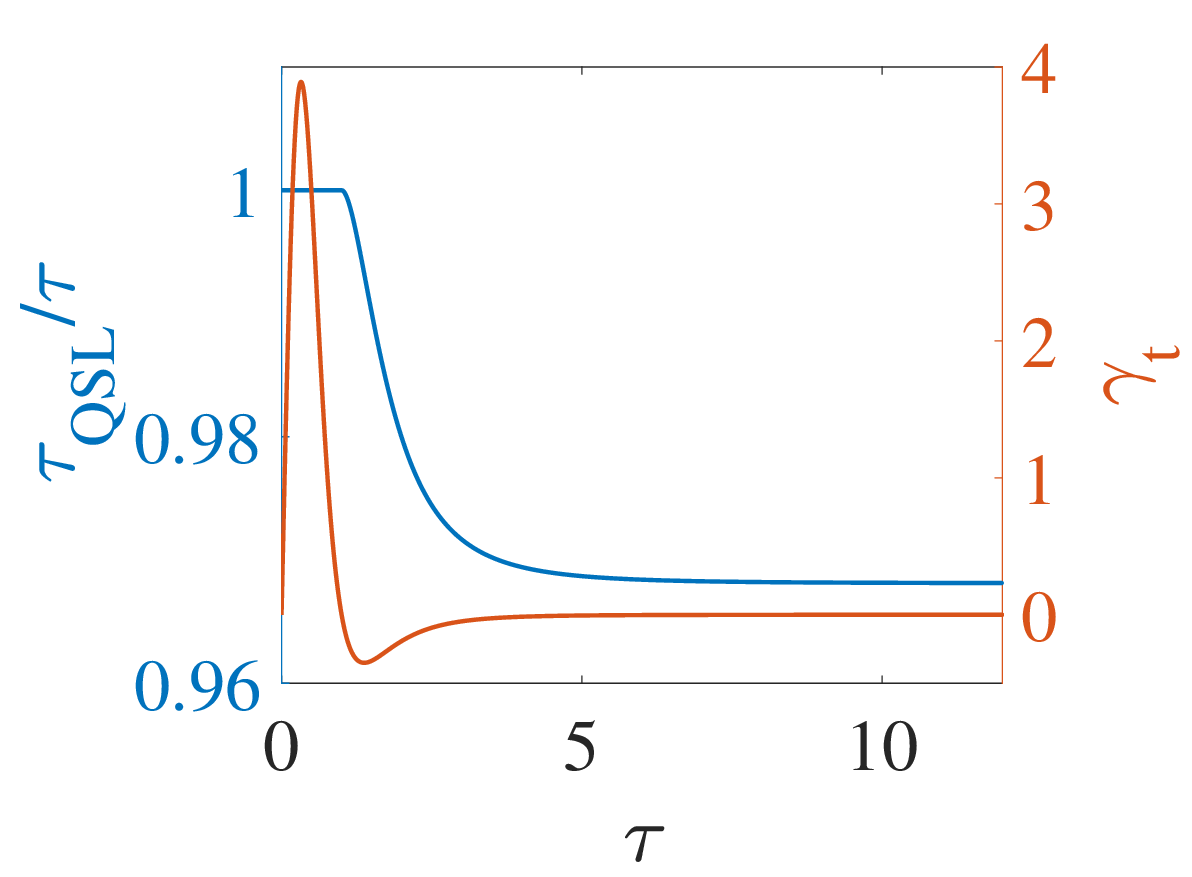}
\caption{(Left)The QSL time $\tau_{\mathrm{QSL}}$ versus the actual unitary evolution time  $\tau$. The red line corresponds to the optimal Hamiltonian $H_T$, and the blue line means $H=H_0+H_1$.  Here $E_m=1$, $\Omega= E_m$ and $\mu=0.5$. (Right)QSL versus evolution time (blue line). The
time-dependent decay rate (red line) is shown as the right axis.  $\protect\omega_c=1$ and $k=4$ are taken for the decay rate in the non-Markovian regime.}
\label{figu}
\end{figure}

As a concrete example,   let us consider a three-level system with the free Hamiltonian $H_0=\mu E_m\left\vert 1\right\rangle\left\langle
1\right\vert+E_m\left\vert 2\right\rangle\left\langle 2\right\vert$, which is initially at the state  $\rho_0=
\left\vert 0\right\rangle\left\langle0\right\vert$.  If an external interaction $H_1=\Omega\left(\left\vert
0\right\rangle\left\langle 2\right\vert+\left\vert
2\right\rangle\left\langle 0\right\vert\right)$ is imposed, and the system will be excited after a duration $\tau$.
Comparably, one can also select the optimal total Hamiltonian as $H_T=U^FH_0(U^F)^\dag$.   It is shown in Fig.  \ref{figu} that the non-optimal Hamiltonian leads to $\tau_{\mathrm{QSL}}$ deviating from the actual evolution time $\tau$, which means the potential of speedup evolution,  but the optimal Hamiltonian $H_T$ indicates that actual evolution time is the same as our bound $\tau_{\mathrm{QSL}}$.

In addition, a beautiful result can also be obtained for the remarkable case of the unitary evolution between two orthogonal pure states as follow:

\textbf{Corolary 4}.-For an $N-$dimensional closed system, the needed time $\tau$ for a given Hamiltonian $H$ to convert an initial pure state $\rho_0$ to its orthogonal state in terms of the distance $\tilde{D}_\alpha$ is lower bounded by
\begin{equation}\label{Uo}
  \tau\geq\tau_{\mathrm{QSL}}=\frac{\frac{2\pi}{3}(N-1)(N^2-1)}{ \sum_{i\neq j}\frac{\sqrt 2}{\sqrt{\mathrm{Tr}\left[\rho_0\right]^2_{ij}-\frac{1}{N}}}\left\langle \Delta E\left(\left[\rho_t\right]_{ij}\right)\right\rangle_\tau},
\end{equation}
where the permutation matrix $U^F$ optimal for the distance of orthogonal states is considered for $[\rho_t]_{ij}$.

The proof of Eq. (\ref{Uo}) can be completed by directly substituting the states and the Hamiltonian into Eqs. (\ref{opt_rho0},\ref{opt_H}) , which is omitted.

One can find that the form of Eq. (\ref{Uo}) is very similar to the remarkable MT bound. A key difference of Eq. (\ref{Uo}) is the dependence on the initial state $\rho_0$, which can be eliminated in the qubit case.

\section{Applications and attainability for open systems}
Due to the particularity of closed systems, we have presented Eq. (\ref{U_QSL}) as a special case of the QSL in Eq. (\ref{QSL2}).  We will turn to Eq. (\ref{QSL2}), the general QSL for open systems.
We will consider a particular non-unitary dynamics that can saturate the inequality of Eq. (\ref{QSL2}).
Thus, we'd like to present the following corollary.

The QSL inequality in Eq. (\ref{QSL2}) with $\alpha_{ij}=\mathrm{Tr}\rho^2_0$  for any N-dimensional initial state $\rho_0$ is saturated by  the depolarizing dynamics
\begin{equation}  \label{open_geod_attain}
\rho_t=p_t\rho_0+(1-p_t)\frac{\mathbb{I}}{N},
\end{equation}
where $p_t$ is a probability function of $t$  and $p_0=1$.  Namely,  the depolarizing dynamics will drive the system to evolve along the geodesic.

According to the definition in Eq. (\ref{map}), it is not difficult to observed that ${F}_\alpha(\rho_t)$  satisfying the form of Eq. (\ref{open_geod_attain}) is just a
linearly composition of $F_\alpha(\rho_0)$ and $F_\alpha(\mathbb{I}/N)$,  and  $\left\langle F_\alpha(\mathbb{I}/N),F_\alpha(\rho)%
\right\rangle=0$ for $\alpha=\mathrm{Tr}\rho_0^2$.  Therefore,
we can define a constant $K=\mathrm{Tr}{\rho_0^2}-1/N$ and construct a  function $\theta_t=\arcsin\left(K(1-p_t)^2/\sqrt{Np_t^2+(1-p_t^2)^2K^2}\right)$ which monotonically depends on $t$ due to the monotonicity of $p_t$,  such that
\begin{equation}  \label{open_F_attain}
\bar{F}_\alpha(\rho_t)=\sin\theta_t
\bar{F}_\alpha\left(\frac{\mathbb{I}}{N}\right)+\cos\theta_t \bar{F}_\alpha(\rho_0).
\end{equation}
 The derivative of $t$ gives
\begin{equation}  \label{open_dF_attain}
\frac{d}{dt}\bar{F}_\alpha(\rho_t)
=\left(\cos\theta_t \bar{F}_\alpha\left(\frac{\mathbb{I}}{N}\right)
-\sin\theta_t \bar{F}_\alpha(\rho_0)
\right)\frac{d}{dt}\theta_t
\end{equation}
and the metric becomes
$
\left\vert\frac{d}{dt}\bar{F}_\alpha(\rho_t)\right\vert^2=\left\vert\frac{d}{dt}%
\theta_t\right\vert^2.
$
Integrating the metric from $t=0$ to $t=\tau$, one can immediately obtain
\begin{eqnarray}  \label{openaa}
&&\int_0^\tau dt\left\vert\frac{d}{dt}\theta_t\right\vert =\left\vert\theta_%
\tau-\theta_0\right\vert=\arccos\left[\cos\left(\theta_\tau-\theta_0\right)%
\right]\notag \\
&&=\arccos\left\langle\bar{F}_\alpha(\rho_\tau),\bar{F}_\alpha(\rho_0)\right\rangle%
=\theta_\alpha(\rho_0,\rho_\tau),
\end{eqnarray}
where the first equality comes from the monotonicity of $\theta_t$.   Thus, QSL in  Eq. (\ref{QSL1}) is saturated.

The eigendecomposition of  $\rho_t$ reads
\begin{equation}  \label{open_rhot_attain}
\rho_t=\Phi\left[p_t\Lambda_0+\left(1-p_t\right)\frac{\mathbb{I}}{N}%
\right]\Phi^\dagger=\Phi\Lambda_t\Phi^\dagger
\end{equation}
with $\Lambda_{t}=p_t\Lambda_0+\left(1-p_t%
\right)\frac{\mathbb{I}}{N}$.  It can be rewritten in the basis $\left\{\Phi\left\vert i_F\right\rangle\right\}$ as
\begin{equation}
{\rho}_t=\Phi\left[\frac{1}{N}\sum_{i=1}^{N}\left\vert i_F\right%
\rangle\left\langle i_F\right\vert+p_t\sum_{i\neq
j}\lambda_{ij}\left\vert i_F\right\rangle\left\langle j_F\right\vert\right]\Phi^\dagger.
\end{equation}
Thus, one can easily obtain the projective matrices as
\begin{equation}  \label{open_rhoij_attain}
\left[\rho_t\right]_{ij}
=p_t\left[\rho_0\right]_{ij}+(1-p_t)\frac{\mathbb{I}}{N}.
\end{equation}
Based on the proof of Eq. (\ref{openaa}),  the QSL  for $\left[\rho_t\right]_{ij}$ is attainable for any $\{i,j\}$ which directly indicates that
the QSL in Eq. (\ref{QSL2}) is saturated by the depolarizing dynamics.

As an application, considering an $N$-level system with Hamiltonian $H_{\mathrm{%
sys}}=\sum_{i=0}^{N-1}E_i\left\vert i\right\rangle\left\langle i\right\vert$
with $E_i$  the energy of the $i$th energy level.  Suppose
the system couples to a heat bath  $H_{\mathrm{B}%
}=\sum_k\omega_k b_k^\dagger b_k$ with $b_k$  the annihilator of the $k$%
th mode of the bath, then the total Hamiltonian reads $H=H_{\mathrm{sys}} +H_{\mathrm{B}}+H_{\mathrm{int}}$
with  $H_{\mathrm{int}}=\sum_{ik}(g_k%
\sigma_+^ib_k+\mathrm{h.c})$ and $\sigma _{-}^{i}=\left\vert 0\right\rangle
\left\langle i\right\vert $ and $\sigma _{+}^{i}=\left\vert i\right\rangle
\left\langle 0\right\vert $ the transition operators.  Following Ref.
\cite{breuer2002theory},  the dynamics can be given as
\begin{equation}
\dot{\rho _{t}}=\sum_ki\left[\rho_t,\frac{s^k_t}{2}\left\vert
k\right\rangle\left\langle k\right\vert\right]+\frac{\gamma _{t}}{2%
}\sum_k\left(2\sigma _{-}^{k}\rho _{t}\sigma _{+}^{k}-\{\sigma _{+}^{k}\sigma
_{-}^{k},\rho _{t}\}\right),\end{equation}
where $s^k_t$ and $\gamma_t$ are the time-dependent Lamb shift and decay
rate, respectively.
The decay rate can take
$
\gamma_t=\int d\omega J(\omega)\coth\left(\frac{\omega}{2k_BT}\right)\frac{%
\sin\omega t}{\omega},
$
where $J(\omega)$ is the Ohmic-like spectral density of the bath. Under the
low-temperature limit, $J(\omega)$ can be expressed as
$
J(\omega)=\frac{\omega^k}{\omega_c^{k-1}}e^{-\frac{\omega}{\omega_c}},
$
where $\omega_c$ denotes the cutoff energy, and  $k$ corresponds to sub-Ohmic ($k<1$), Ohmic ($k=1$) and super-Ohmic ($k>1$)-type
environment. Under
the zero temperature limit, the decay rate reads.
$
\gamma_t=\omega_c\left(1+\omega_c^2t^2\right)^{-k/2}\Gamma\left(k\right)\sin
\left[k\arctan\left(\omega_c t\right)\right],
$
where $\Gamma$ is the Euler function.

\begin{figure}
\includegraphics[width=0.48\linewidth]{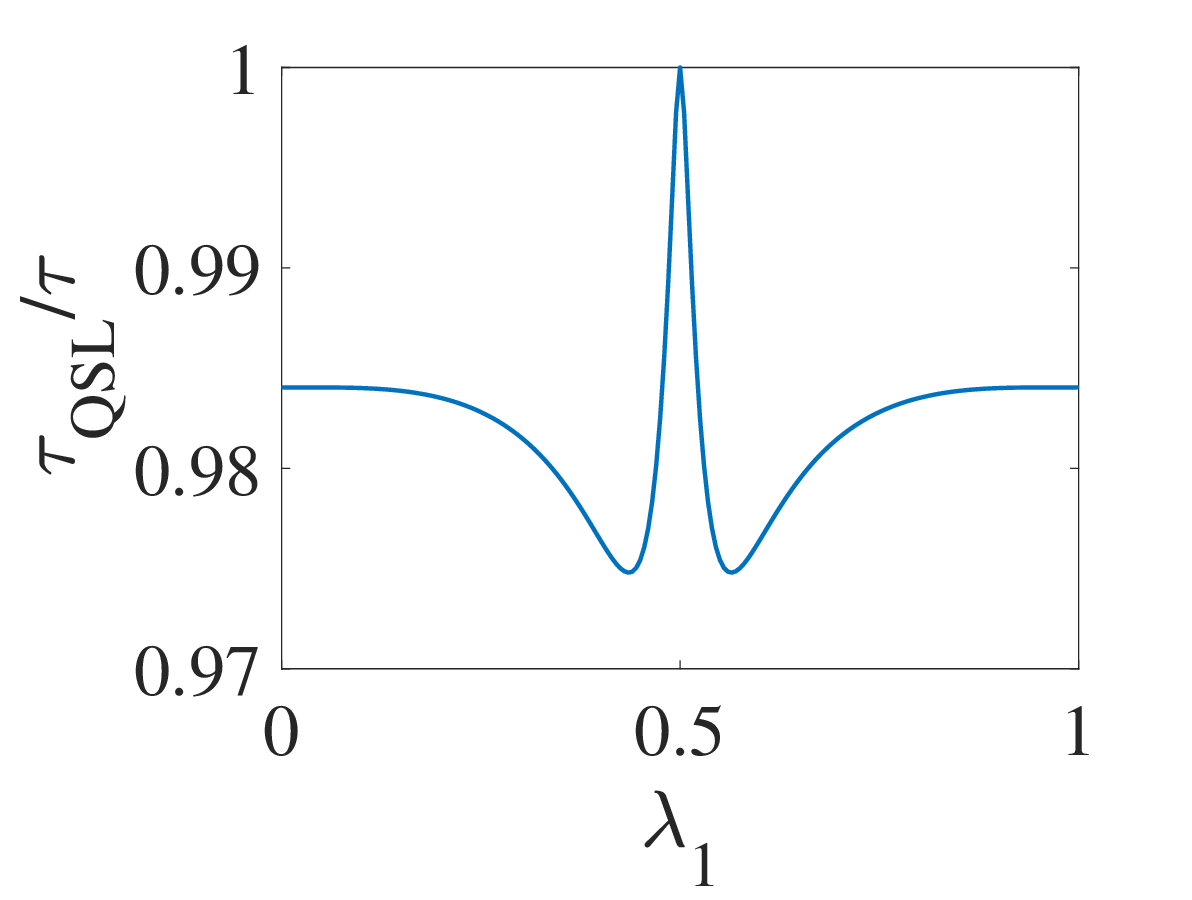}%
\includegraphics[width=0.48\linewidth]{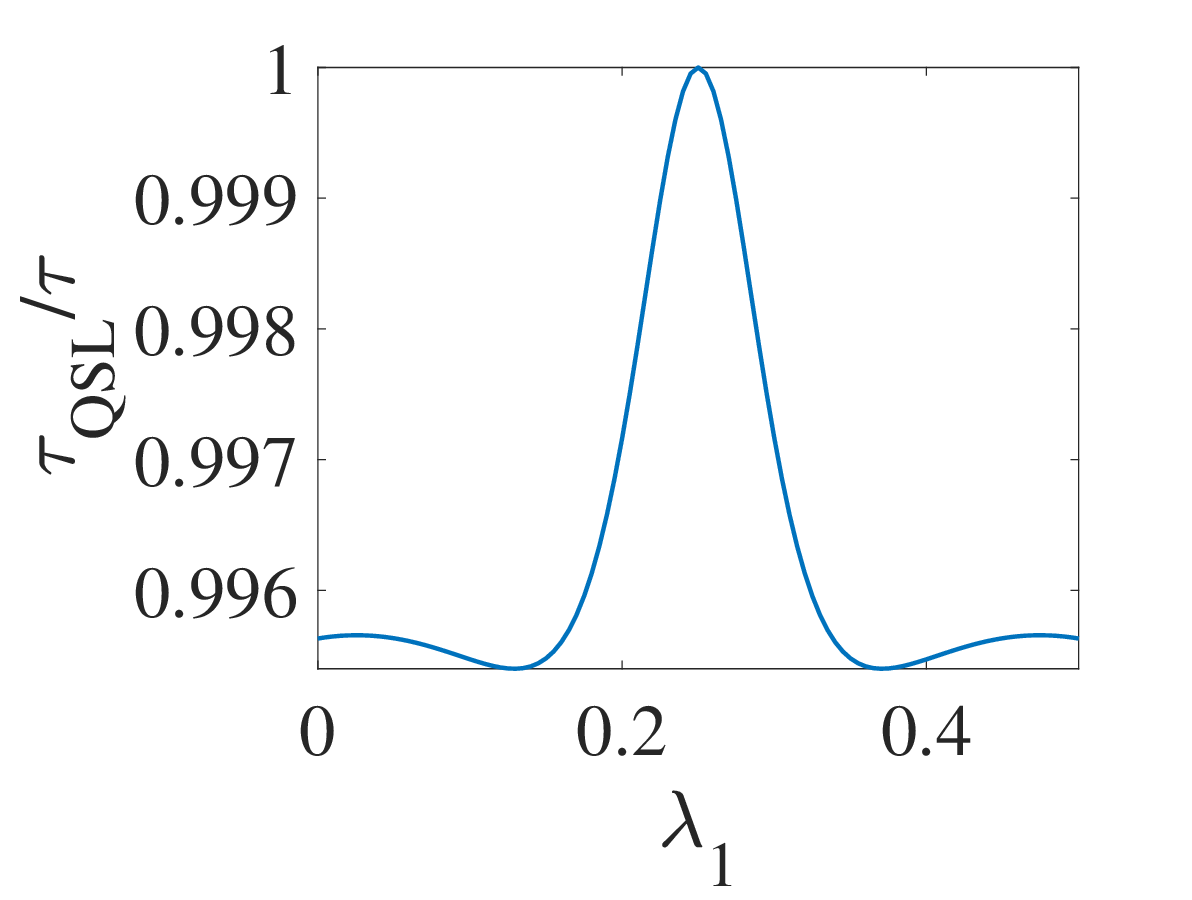}
\caption{Different initial state vs corresponding QSL. The evolution time is
$\protect\tau=1$. The system is prepared as excited state initially with $%
\protect\lambda_0=0$ (Left), and the ground state is partly occupied with $%
\protect\lambda_0=0.5$ (Right), respectively.}\label{jcf}
\end{figure}

Take a $3$-dimensional initial state $%
\rho_0=\sum_{i=0}^2\lambda_i\left\vert i\right\rangle\left\langle
i\right\vert$ as an example. The state at the instant $t$ reads
\begin{equation}
\rho_t=\left(1-e^{-\Gamma_t}\sum_{i=1}^2\lambda_i\right)\left\vert
0\right\rangle\left\langle
0\right\vert+e^{-\Gamma_t}\sum_{i=1}^2\lambda_i\left\vert
i\right\rangle\left\langle i\right\vert \label{jc_solu}
\end{equation}
with $\Gamma_t=\int_0^t dt^{\prime }\gamma_{t^{\prime }}$.
To show the non-Markovian effect  on the QSL, we let a maximally mixed state undergo the above dynamics and numerically calculate the
QSL with different evolution times under the non-Markovian regime in Fig \ref{figu} (Right). The QSL is saturated with tiny evolution time $\tau$ when
$\gamma_t$ is positive.   With $\tau$ increasing, $\gamma_t$ rapidly
decreases to the negative value,  which means the information flows back from
the environment. Hence, it indicates that the non-Markovian effect breaks the
saturation of the QSL, which is similar to Ref. \cite%
{deffner2013quantum,PhysRevA.93.020105,PhysRevA.96.012105}.
Considering Markovian dynamics, i.e.,  $\gamma_t=1$,  one can get the QSL with different initial states as shown in Fig. \ref{jcf}. It
can be seen in Fig. \ref{jcf} that our QSL is saturated for $%
\lambda_1=\lambda_2$ in which case Eq. (\ref{jc_solu}) can be written as the form of Eq. (\ref{open_rhot_attain}).

\section{conclusion and discussion}
We establish new QSL bounds that are valid for open
and closed systems.  It indicates that for any density matrix, one can always find unitary
and non-unitary dynamics to saturate their corresponding QSLs in any dimensional system, and given a Hamiltonian, one can also find a pair of states to saturate the QSL bound. Thus, we have solved the long-standing problem of QSL attainability. As applications, we have also studied the QSL times in various physical systems. We believe that the current work will shed new light on QSL research.

\section{Acknowledgments}
This work was supported by the National Natural Science Foundation of China under Grant No. 12175029.

\bibliography{reference}


\end{document}


\title{Supplementary Information for\\
Attainable quantum speed limit for N-dimensional quantum systems}

\maketitle
\onecolumngrid

\tableofcontents
\beginsupplement


\section{$D_\alpha$ is a good distance for quantum states}
We will show that the distance given in Definition 1 is good for quantum states.  It is obvious that $D_\alpha\geq 0$ holds due to the Cauchy-Schwarz inequality,  i.e.,
\begin{equation}  \label{appc_csineq}
D_\alpha\left(\rho,\sigma\right)=\arccos\left\langle \frac{%
F_\alpha(\rho)}{\left\vert F_\alpha(\rho)\right\vert}, \frac{F_\alpha(\sigma)%
}{\left\vert F_\alpha(\sigma)\right\vert}\right\rangle\geq 0,
\end{equation}
where the equality holds iff
\begin{equation}  \label{appc_csineq2}
\frac{F_\alpha(\rho)}{\left\vert F_\alpha(\rho)\right\vert}=\frac{%
F_\alpha(\sigma)}{\left\vert F_\alpha(\sigma)\right\vert}.
\end{equation}

 We will show that Eq. (\ref{appc_csineq2}) equals $\rho=\sigma$.  It
is obvious that Eq. (\ref{appc_csineq2}) holds when $\rho=\sigma$,  so we will show
Eq. (\ref{appc_csineq2}) can lead to  $\rho=\sigma$.  Let's trace over both sides of  Eq. (\ref{appc_csineq2}), one can immediately
obtain
\begin{equation}  \label{appc_trace}
\frac{\mathrm{Tr}\rho^2-\alpha}{\sqrt{\mathrm{Tr}\rho^2-\frac{1}{N}+\frac{1}{%
N}\left(\alpha-\mathrm{Tr}\rho^2\right)^2}} =\frac{\mathrm{Tr}\sigma^2-\alpha%
}{\sqrt{\mathrm{Tr}\sigma^2-\frac{1}{N}+\frac{1}{N}\left(\alpha-\mathrm{Tr}%
\sigma^2\right)^2}}.
\end{equation}
Denote $x=\mathrm{Tr}\rho^2-1/N$, $y=\mathrm{%
Tr}\sigma^2-1/N$ and $z=\alpha-1/N$, one can find that
\begin{equation}  \label{appc_trace_2}
\frac{(x-z)^2}{x+\frac{1}{N}(x-z)^2}=\frac{(y-z)^2}{y+\frac{1}{N}(y-z)^2}.
\end{equation}

If $x=z$, Eq. (\ref{appc_trace_2}) means $y=z$, that is, $\mathrm{Tr}%
\rho^2=\mathrm{Tr}\sigma^2$.  If  $y=z$,  one will obtain the same result.

 If $x\neq z$ and $%
y\neq z$, Eq. (\ref{appc_trace_2}) means
\begin{equation}  \label{appc_trace_3}
\frac{x}{(x-z)^2}=\frac{y}{(y-z)^2},
\end{equation}
if $xy=0$, we have $\mathrm{Tr}\rho^2=\mathrm{Tr}\sigma^2=1/N$. If $%
x\neq 0$ and $y\neq 0$, Eq. (\ref{appc_trace_3}) can be further expressed as
\begin{equation}  \label{appc_trace_4}
x-y=z^2\frac{x-y}{xy},
\end{equation}
which means that $x=y$ or $xy=z^2$.  Substituting $y=z^2/x$ into Eq. (\ref%
{appc_trace}), one has $(x-z)/\sqrt{x+(x-z)^2/N}=(z-y)/\sqrt{y+(y-z)^2/N}$%
, which leads to $x=y=z$. To sum up,  Eq. (\ref{appc_trace}) means $\mathrm{Tr}\rho^2=\mathrm{Tr}\sigma^2$ which leads to $\left\vert
F(\rho)\right\vert=\left\vert F(\sigma)\right\vert$. Thus, one will arrive at $%
\rho=\sigma$ according to  Eq. (\ref{appc_csineq2}).

Now, we will prove $D_\alpha$ satisfies the triangular inequality.  First, we will show that for any Hermitian matrix $A=A^\dagger\neq 0$ and $B=B^\dagger\neq 0$, $d(A,B)=\arccos\frac{\left\langle A,B\right\rangle}{\left\vert A\right\vert\cdot\left\vert B\right\vert}$ is the Euclidean distance.  Suppose
$
  A=\sum_{j,k} a_{jk}\left\vert j\right\rangle\left\langle k\right\vert
$
with $a_{jk}=x^A_{jk}+iy_{jk}^A$, $x_{jk}^A,y_{jk}^A\in\mathbb R$.  It is not difficult to calculate that $\left\langle A, B\right\rangle=\sum_{j,k}\left(x_{jk}^Ax_{jk}^B+y_{jk}^Ay_{jk}^B\right)=\langle A\vert B\rangle$, where
\begin{equation}\label{appe_2}
  \vert A\rangle=\begin{pmatrix}
                   \vert x^A\rangle \\
                   \vert y^A\rangle
                 \end{pmatrix},
  \vert x^A\rangle=\begin{pmatrix}
                     x_{11}^A \\
                     x_{12}^A \\
                     ... \\
                     x_{NN}^A
                   \end{pmatrix},   \vert y^A\rangle=\begin{pmatrix}
                     y_{11}^A \\
                     y_{12}^A \\
                     ... \\
                     y_{NN}^A
                   \end{pmatrix}
\end{equation}
and $\vert A\rangle$ is defined similarly. Hence, $\langle A, B\rangle$ is the inner product of two vectors $\vert A\rangle$ and $\vert B\rangle$,  which indicates that $d(A, B)$ is the angle of two points onto the unit sphere within the Euclidean space, which satisfies the triangle inequality.
In other words, $D_\alpha$ for $%
F(\rho)\neq 0$ and $%
F(\sigma)\neq 0$ satisfies the triangular inequality, too.

Next we will show that  $F(\rho)=0$ is impossible for density
matrix $\rho$.  We assume that $F(\rho)=0$ which directly leads to
\begin{equation}  \label{appc_triangle}
\rho=\left(1+\alpha-\mathrm{Tr}\rho^2\right)\mathbb{I}/N.
\end{equation}
Square both sides of  Eq. (\ref{appc_triangle}) and then take
trace, we have
\begin{equation}  \label{appc_triangle2}
\mathrm{Tr}\rho^2=\frac{\left(\alpha-\mathrm{Tr}\rho^2\right)^2}{N}.
\end{equation}
It can be solved that
\begin{equation}  \label{appc_triangle3}
\mathrm{Tr}\rho^2=\frac{N}{2}+1+\alpha\pm\frac{\sqrt{N^2+4N(1+\alpha)}}{2}.\end{equation}
 Substituting Eq. (\ref{appc_triangle3}) into Eq. (\ref{appc_triangle}%
) and tracing both sides, one will arrive at
\begin{eqnarray}  \label{appc_triangle4}
1=&\mathrm{Tr}\rho=\mp\frac{\sqrt{N^2+4N(1+\alpha)}}{2}-\frac{N}{2} \notag\\
=&\frac{\sqrt{N^2+4N(1+\alpha)}}{2}-\frac{N}{2} \notag\\
>&\frac{\sqrt{N^2+4N\left(1+\frac{1}{N}\right)}}{2}-\frac{N}{2}=1,
\end{eqnarray}
where the final inequality comes from the condition $\alpha>1/N$,  which  indicates a contradiction.  So Eq. (\ref{appc_triangle}) is impossible. To sum up, we can safely say that the distance  $D_\alpha $ satisfies the triangle inequality.

\section {Derivation of  QSL time in different systems}

Consider a 3-dimensional case.  Suppose $\rho_0$ is the intial state fixed as the diagonal form as
\begin{equation}\label{rho0}
  \rho_0=\begin{pmatrix}
           \lambda_0 & 0 & 0 \\
           0 & \lambda_1 & 0 \\
           0 & 0 & \lambda_2
         \end{pmatrix},
\end{equation}
where $\lambda_0=1-(\lambda_1+\lambda_2)$ with positive $\lambda_{0,1,2}$.
We introduce the Flourier transformation as
\begin{equation}\label{U_F}
  \left\langle m\right\vert U_F\left\vert n\right\rangle=\frac{1}{\sqrt 3}e^{2i\pi\frac{(m+1)(n+1)}{3}}
\end{equation}
for $m,n=0,1,...,N-1$. For example, the $3\times 3$ Flourier transformation can be expressed as
\begin{equation}\label{U_F3}
  U_F=\frac{1}{\sqrt 3}\begin{pmatrix}
                         e^{i\pi\frac{2}{3}} & e^{i\pi\frac{4}{3}} & 1 \\
                         e^{i\pi\frac{4}{3}} & e^{i\pi\frac{2}{3}} & 1 \\
                         1 & 1 & 1
                       \end{pmatrix}.
\end{equation}
Then, one can calculate the projective matrices  as
\begin{equation}\label{uni_sub_matrix}
\begin{split}
  [\rho_0]_{01} &= \frac{\mathbb I}{3}+\frac{1}{3}\begin{pmatrix}
                                                     \tilde\lambda_0 & \tilde\lambda_2 & \tilde\lambda_1 \\
                                                     \tilde\lambda_2 & \tilde\lambda_1 & \tilde\lambda_0 \\
                                                     \tilde\lambda_1 & \tilde\lambda_0 & \tilde\lambda_2
                                                   \end{pmatrix} \\
 [\rho_0]_{02} &= \frac{\mathbb I}{3}+\frac{1}{3}\begin{pmatrix}
                                                     \tilde\lambda_0 & e^{i\frac{2}{3}\pi}\tilde\lambda_2 & -e^{i\frac{\pi}{3}}\tilde\lambda_1 \\
                                                     e^{-i\frac{2}{3}\pi}\tilde\lambda_2 & \tilde\lambda_1 & -e^{-i\frac{\pi}{3}}\tilde\lambda_0 \\
                                                     -e^{-i\frac{\pi}{3}}\tilde\lambda_1 & -e^{i\frac{\pi}{3}}\tilde\lambda_0 & \tilde\lambda_2
                                                   \end{pmatrix} \\
 [\rho_0]_{12} &= \frac{\mathbb I}{3}+\frac{1}{3}\begin{pmatrix}
                                                     \tilde\lambda_0 & e^{-i\frac{2}{3}\pi}\tilde\lambda_2 & -e^{-i\frac{\pi}{3}}\tilde\lambda_1 \\
                                                     e^{i\frac{2}{3}\pi}\tilde\lambda_2 & \tilde\lambda_1 & -e^{i\frac{\pi}{3}}\tilde\lambda_0 \\
                                                     -e^{i\frac{\pi}{3}}\tilde\lambda_1 & -e^{-i\frac{\pi}{3}}\tilde\lambda_0 & \tilde\lambda_2
                                                   \end{pmatrix}
\end{split}
\end{equation}
with $\tilde\lambda_i=\lambda_i-\frac{1}{3}$.

\subsection{The unitary case}
In the closed system, let's consider two unitary evolutions governed by two Hamiltonians, respectively, given as
\begin{eqnarray}
  H=&\begin{pmatrix}
      0 & 0 & \Omega \\
      0 & \mu E_m & 0 \\
      \Omega & 0 & E_m
    \end{pmatrix}\label{H_unitary}\\
  H_{\mathrm{opt}}=&U_F^\dagger\begin{pmatrix}
                       0 & 0 & 0 \\
                       0 & \mu E_m & 0 \\
                       0 & 0 & E_m
                     \end{pmatrix}U_F\label{H_opt},
\end{eqnarray}
where $H_{\mathrm{opt}}$ saturates our QSL.

According to the Hamiltonian $H$, the corresponding evolution operator is
\begin{eqnarray}\label{Ut}
  U_t=&\begin{pmatrix}
        0 & \frac{\Delta}{\sqrt{\Delta^2+4\Omega^2}} & -\frac{2\Omega}{\sqrt{\Delta^2+4\Omega^2}} \\
        1 & 0 & 0 \\
         0& \frac{2\Omega}{\sqrt{\Delta^2+4\Omega^2}} & \frac{\Delta}{\sqrt{\Delta^2+4\Omega^2}}
      \end{pmatrix}
      \begin{pmatrix}
           e^{-i\mu E_mt} & 0 & 0 \\
           0 & e^{-i\frac{2E_m+\Delta}{2}t} & 0 \\
           0 & 0 & e^{i\frac{\Delta}{2}t}
         \end{pmatrix}
      &\begin{pmatrix}
        0 & 1 & 0 \\
        \frac{\Delta}{\sqrt{\Delta^2+4\Omega^2}} & 0 & \frac{2\Omega}{\sqrt{\Delta^2+4\Omega^2}} \\
        -\frac{2\Omega}{\sqrt{\Delta^2+4\Omega^2}} & 0 & \frac{\Delta}{\sqrt{\Delta^2+4\Omega^2}}
      \end{pmatrix}.
\end{eqnarray}
Thus the projective matrices at the time $t$ can be given as $U_t[\rho_0]_{ij}U_t^\dagger$.  Then one can calculate $\Delta E$ of the projective matrices as
\begin{eqnarray}\label{unitary_metric}
  \Delta E\left(\left[\rho_0\right]_{01}\right)=&&\frac{1}{3}\Big\{\left[(\mu-1)^2E_m^2+2\Omega^2\right]\tilde\lambda_0^2+E_m^2\tilde\lambda_1^2+
  \left(\mu^2E_m^2+2\Omega^2\right)\tilde\lambda_2^2+2 E_m\Omega\tilde\lambda_0\tilde\lambda_1+\notag\\
&&  \left(2\Omega^2+E_m\Omega-2\mu E_m\Omega\right)\tilde\lambda_0\tilde\lambda_2-2 E_m\Omega\tilde\lambda_1\tilde\lambda_2\Big\},\notag\\
  \Delta E\left(\left[\rho_0\right]_{02}\right)=&&\frac{1}{3}\Big\{\left[(\mu-1)^2E_m^2+2\Omega^2\right]\tilde\lambda_0^2+\left(E_m^2+3\Omega^2\right)\tilde\lambda_1^2+
  \left(\mu^2E_m^2+2\Omega^2\right)\tilde\lambda_2^2- E_m\Omega\tilde\lambda_0\tilde\lambda_1-\\
&&\left(2\Omega^2+E_m\Omega-2\mu E_m\Omega\right)\tilde\lambda_0\tilde\lambda_2+ E_m\Omega\tilde\lambda_1\tilde\lambda_2\Big\},\notag\\
  \Delta E\left(\left[\rho_0\right]_{12}\right)=&&\Delta E\left(\left[\rho_0\right]_{02}\right).\notag
\end{eqnarray}
Substituting Eqs. (\ref{uni_sub_matrix},\ref{Ut},\ref{unitary_metric}) into Corollary 3,  one can obtain the QSL times under different evolution times $\tau$.  For simplicity, we only consider the unitary evolution from the ground initial state $\rho_0=\left\vert 0\right\rangle\left\langle 0\right\vert$, the QSL time compared with the saturation case is numerically shown in Fig. 1 in the main text.

\subsection{The dissipative case}
Similar to Eq. (\ref{uni_sub_matrix}), if the initial state is prepared as the diagonal form,  it is not difficult to obtain the projective matrices as
\begin{eqnarray} \label{jc_sub_matrix}
  [\rho_t]_{01} &=& \frac{\mathbb I}{3}+\frac{1}{3}\begin{pmatrix}
                                                     \tilde\lambda_0(t) & \tilde\lambda_2(t) & \tilde\lambda_1(t) \\
                                                     \tilde\lambda_2(t) & \tilde\lambda_1(t) & \tilde\lambda_0(t) \\
                                                     \tilde\lambda_1(t) & \tilde\lambda_0(t) & \tilde\lambda_2(t)
                                                   \end{pmatrix}      \end{eqnarray}
      \begin{eqnarray}
  [\rho_t]_{02} &= &\frac{\mathbb I}{3}+\frac{1}{3}\begin{pmatrix}
                                                     \tilde\lambda_0(t) & e^{i\frac{2}{3}\pi}\tilde\lambda_2(t) & -e^{i\frac{\pi}{3}}\tilde\lambda_1(t) \\
                                                     e^{-i\frac{2}{3}\pi}\tilde\lambda_2(t) & \tilde\lambda_1(t) & -e^{-i\frac{\pi}{3}}\tilde\lambda_0(t) \\
                                                     -e^{-i\frac{\pi}{3}}\tilde\lambda_1(t) & -e^{i\frac{\pi}{3}}\tilde\lambda_0(t) & \tilde\lambda_2(t)
                                                   \end{pmatrix}  \end{eqnarray}
     \begin{eqnarray}
    [\rho_t]_{12} &= &\frac{\mathbb I}{3}+\frac{1}{3}\begin{pmatrix}
                                                     \tilde\lambda_0(t) & e^{-i\frac{2}{3}\pi}\tilde\lambda_2(t) & -e^{-i\frac{\pi}{3}}\tilde\lambda_1(t) \\
                                                     e^{i\frac{2}{3}\pi}\tilde\lambda_2(t) & \tilde\lambda_1(t) & -e^{i\frac{\pi}{3}}\tilde\lambda_0(t) \\
                                                     -e^{i\frac{\pi}{3}}\tilde\lambda_1(t) & -e^{-i\frac{\pi}{3}}\tilde\lambda_0(t) & \tilde\lambda_2(t)
                                                   \end{pmatrix}
\end{eqnarray}
with $\tilde\lambda_{1,2}(t)=p_t\tilde\lambda_{1,2}-\frac{1}{3}(1-p_t)$ and $\tilde\lambda_{0}(t)=-\left[\tilde\lambda_{1}(t)+\tilde\lambda_{2}(t)\right]$. Meanwhile, it can be observed that
\begin{equation}\label{jc_sub_matrix_rhotij}
  \left[\rho_t\right]_{ij}=p_t\left[\rho_0\right]_{ij}+(1-p_t)\left[\left\vert 0\right\rangle\left\langle 0\right\vert\right]_{ij}
\end{equation}
for $\forall i,j=1,2,3, i<j$, where $\left[\left\vert 0\right\rangle\left\langle 0\right\vert\right]_{ij}$ can be obtained from Eq. (\ref{uni_sub_matrix}) by letting $\lambda_0=1$. Especially, when $\lambda_1=\lambda_2=\lambda$, we have

\begin{equation}\label{jc_sub_matrix_equal}
  \left[\rho_t\right]_{ij}=\frac{3p_t\lambda-1}{3\lambda-1}\left[\rho_0\right]_{ij}+3\lambda\left(p_t-\frac{3p_t\lambda-1}{3\lambda-1}\right)\frac{\mathbb I}{3},
\end{equation}
which is the case  saturating our QSL. To calculate the QSL,  one will have to calculate the following expressions:
\begin{eqnarray}\label{jc_tr}
     &\mathrm{Tr}\left[\rho_t\right]_{ij}^2  =\frac{5}{9}+\frac{2}{3}\left[p_t^2(\lambda_1^2+\lambda_2^2+\lambda_1\lambda_2)-p_t(\lambda_1+\lambda_2)\right], \\
     &\mathrm{Tr}\left(\frac{d}{dt}\left[\rho_t\right]_{ij}\right)^2  =\frac{2}{3}\dot p_t^2(\lambda_1^2+\lambda_2^2+\lambda_1\lambda_2) ,\\
      &\mathrm{Tr}\left(\left[\rho_t\right]_{ij}\frac{d}{dt}\left[\rho_t\right]_{ij}\right) =\frac{1}{3}\dot p_t\left[2p_t(\lambda_1^2+\lambda_2^2+\lambda_1\lambda_2)-(\lambda_1+\lambda_2)\right].
\end{eqnarray}
Substituting all the above expressions into the QSL in Theorem 1, one can obtain the QSL time,   which is provided in a numercial way with different cases in Fig. 2.

\section{QSL time between two orthogonal pure states}
In the main text, we have mentioned that distance could depend on the permutation of eigenvalues, so we take a maximum to eliminate the influence. Here, we would like to demonstrate the effect.
Let's consider the distance of a pair of orthogonal pure states.  Let $\rho=U\vert m\rangle\langle m\vert U^\dagger$ and $\sigma=U\vert n\rangle\langle n\vert U^\dagger$, then the projective matrices can be given as
\begin{eqnarray}\label{rho_pure}
 [ \rho]_{ij}=&UU_P^\dagger U_F^\dagger\left[ P_{ij}U_F\left( U_P\vert m\rangle\langle m\vert U_P^\dagger\right)U_F^\dagger P_{ij}+\frac{1}{N}\left(\mathbb I-P_{ij}\right)\right]U_FU_PU^\dagger\notag\\
  \equiv&UU_P^\dagger U_F^\dagger\left[ P_{ij}U_F\left( \vert \tilde m\rangle\langle \tilde m\vert \right)U_F^\dagger P_{ij}+\frac{1}{N}\left(\mathbb I-P_{ij}\right)\right]U_FU_PU^\dagger,\\
 \left[\sigma\right]_{ij}=&UU_P^\dagger U_F^\dagger\left[ P_{ij}U_F\left( U_P\vert n\rangle\langle n\vert U_P^\dagger\right)U_F^\dagger P_{ij}+\frac{1}{N}\left(\mathbb I-P_{ij}\right)\right]U_FU_PU^\dagger\notag\\
  \equiv&UU_P^\dagger U_F^\dagger\left[ P_{ij}U_F\left( \vert \tilde n\rangle\langle \tilde n\vert\right)U_F^\dagger P_{ij}+\frac{1}{N}\left(\mathbb I-P_{ij}\right)\right]U_FU_PU^\dagger,
\end{eqnarray}
where $U_F$ denotes the Fourier transformation, $U_P$ denotes the permutation operation and $\vert \tilde m(\tilde n)\rangle=U_P\vert m(n)\rangle$. A simple algebra shows that
\begin{eqnarray}\label{rho_pure_project}
  P_{ij}U_F\vert \tilde m\rangle\langle \tilde m\vert U_F^\dagger P_{ji}+\frac{1}{N}(\mathbb I-P_{ij})=\frac{\mathbb I}{N}+\frac{1}{N}e^{2\pi i}\frac{\tilde m}{N}(i-j)\vert i\rangle\langle j\vert+\mathrm{h.c.},\\
  P_{ij}U_F\vert \tilde n\rangle\langle \tilde n\vert U_F^\dagger P_{ij}+\frac{1}{N}(\mathbb I-P_{ij})=\frac{\mathbb I}{N}+\frac{1}{N}e^{2\pi i}\frac{\tilde n}{N}(i-j)\vert i\rangle\langle j\vert+\mathrm{h.c.},
\end{eqnarray}
then it can be calculated that
\begin{eqnarray}\label{rho_pure_tr}
  \mathrm{Tr}[\rho]_{ij}[\sigma]_{ij}&=&\mathrm{Tr}\left[P_{ij}U_F\vert \tilde m\rangle\langle \tilde m\vert U_F^\dagger P_{ji}+\frac{1}{N}(\mathbb I-P_{ij})\right]\left[P_{ij}U_F\vert \tilde n\rangle\langle \tilde n\vert U_F^\dagger P_{ij}+\frac{1}{N}(\mathbb I-P_{ij})\right]\notag\\
 & =&\frac{1}{N}+\frac{1}{N^2}\mathrm{Tr}\begin{pmatrix}
                                          0 & e^{2\pi i\frac{\tilde m}{N}}(i-j) \\
                                          e^{-2\pi i\frac{\tilde m}{N}}(i-j) & 0
                                        \end{pmatrix}\begin{pmatrix}
                                                       0 & e^{2\pi i\frac{\tilde n}{N}}(i-j) \\
                                                       e^{-2\pi i\frac{\tilde n}{N}}(i-j) & 0
                                                     \end{pmatrix}\notag\\
  &=&\frac{1}{N}+\frac{1}{N^2}\mathrm{Tr}\begin{pmatrix}
                                          e^{2\pi i\frac{(\tilde m-\tilde n)(i-j)}{N}} & 0 \\
                                          0 & e^{-2\pi i\frac{(\tilde m-\tilde n)(i-j)}{N}}
                                        \end{pmatrix}\notag\\
&  =&\frac{1}{N}+\frac{1}{N^2} 2\mathrm{Re}\left\{e^{2\pi i\frac{(\tilde m-\tilde n)(i-j)}{N}}\right\}\notag\\
 & =&\frac{1}{N}+\frac{2}{N^2}\cos\left[\frac{(\tilde m-\tilde n)(i-j)}{N}2\pi\right],\\
  \mathrm{Tr}[\rho]_{ij}^2&=&\frac{1}{N}+\frac{2}{N^2}.
\end{eqnarray}
Substituting Eq. (\ref{rho_pure_tr}) into the distance $\tilde{D}_{\alpha}^{\Phi,\Psi}$ with $\Phi$  and $\Psi$ denoting the eigenvector matrices of the initial states, and taking $\alpha_{ij}=\mathrm{Tr}[\rho]_{ij}^2$, we have
\begin{equation}\label{dis_rhoij_pure}
  \tilde{D}_{\alpha_{ij}}^{\Phi,\Psi}([\rho]_{ij},[\sigma]_{ij})=\arccos\frac{\mathrm{Tr}[\rho]_{ij}[\sigma]_{ij}-\frac{1}{N}}{\mathrm{Tr}[\rho]_{ij}^2-\frac{1}{N}}=\left\vert \frac{(\tilde m-\tilde n)(i-j)}{N}\right\vert 2\pi.
\end{equation}
 Summing over all  $i\neq j$, we have $\sum_{i\neq j}\vert i-j\vert=\frac{1}{3}N(N^2-1)$, then the distance reads
 \begin{equation}\label{dis_rho_pure}
  \tilde{D}_{\alpha_{ij}}^{\Phi,\Psi}([\rho]_{ij},[\sigma]_{ij})=\left\vert\tilde m-\tilde n\right\vert\frac{2\pi}{3}\left(N^2-1\right),
 \end{equation}
Finally, optimizing over all the  permutation matrices $U_P$  to maximize   $\vert \tilde m-\tilde n\vert$ gives
\begin{equation}\label{Dis_rho_pure}
  \tilde{D}_{\alpha}^{\Phi,\Psi}([\rho]_{ij},[\sigma]_{ij})=\frac{2\pi}{3}(N-1)(N^2-1).
\end{equation}
Based on Eq. (\ref{Dis_rho_pure}),  one will easily find the QSL time in Corollary 5.

\section{Application to the QSL $\tau_\alpha$}
In the main text, we have presented a QSL in Theorem 1, and we discussed no more about it. In this section, we will apply this QSL $\tau_\alpha$ to some concrete dynamics case.

\begin{figure}
\includegraphics[width=0.38\linewidth]{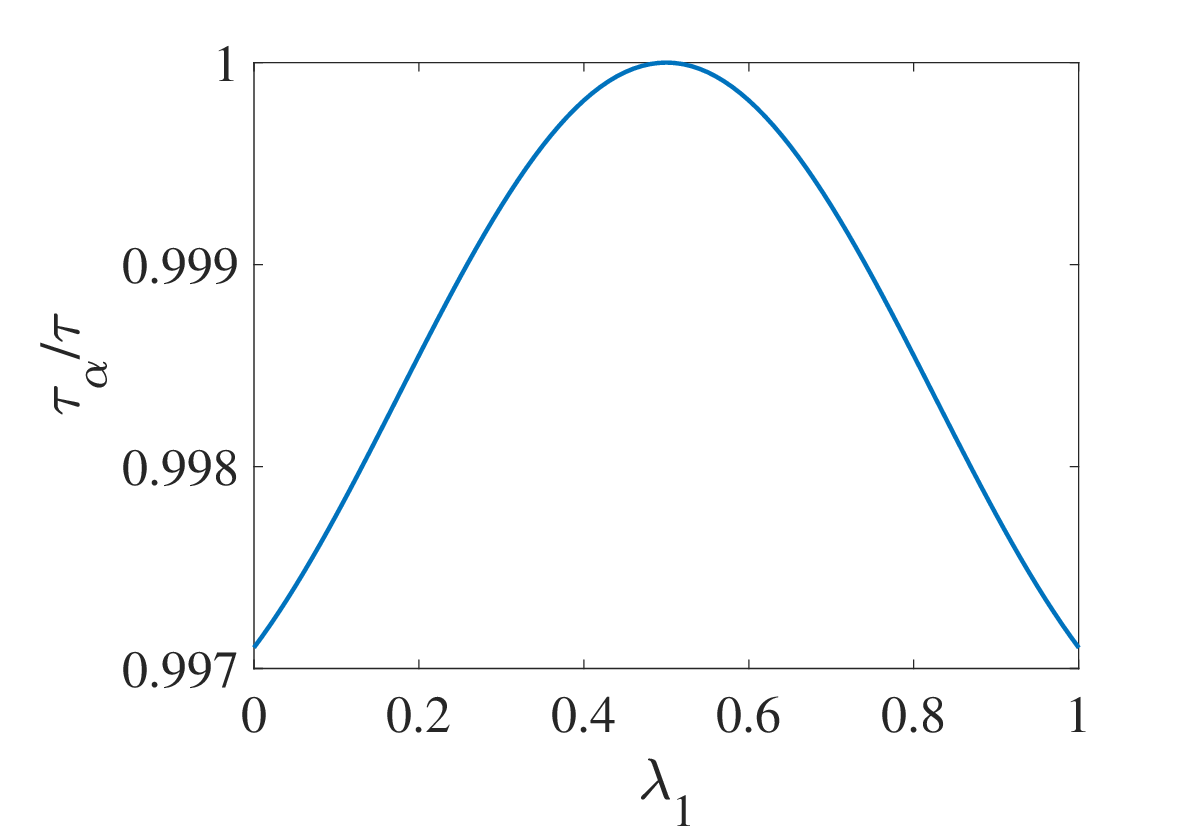}%
\includegraphics[width=0.38\linewidth]{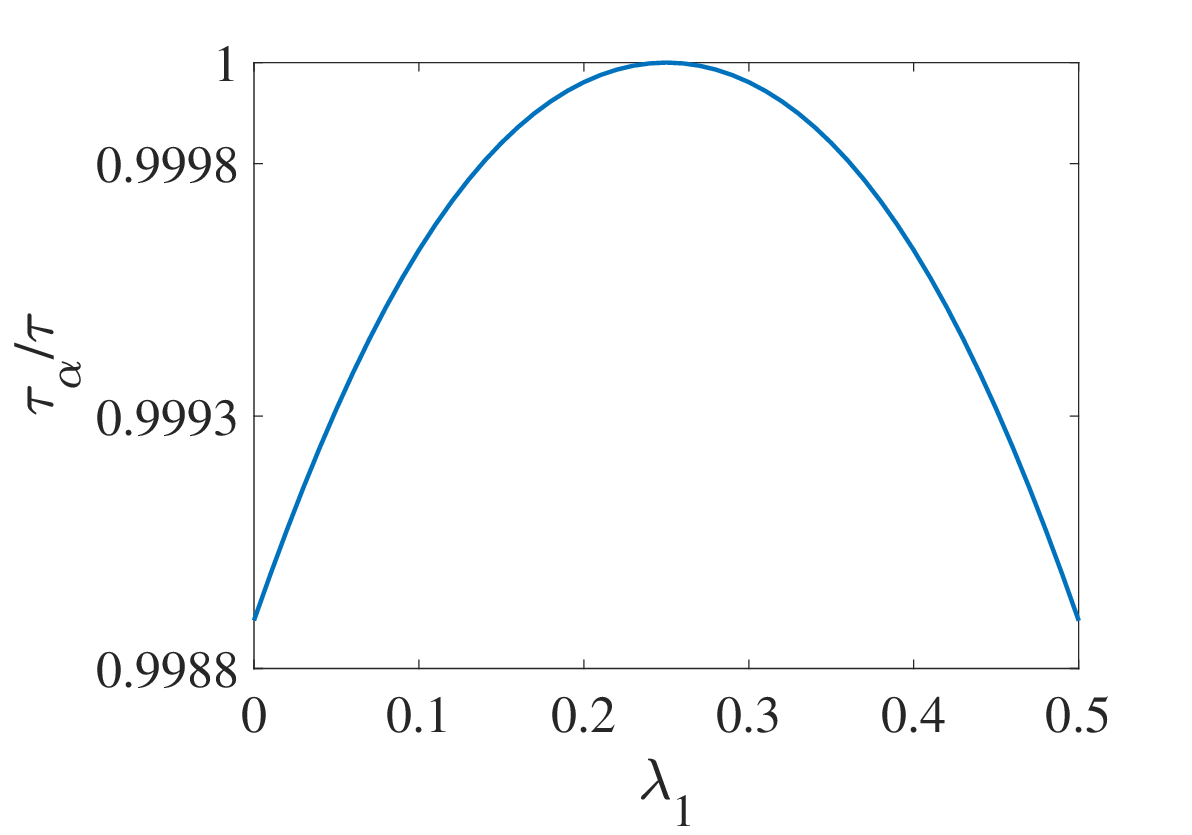}
\caption{Different initial state vs corresponding QSL. The evolution time is
$\protect\tau=1$. The system is prepared as excited state initially with $%
\protect\lambda_0=0$ (Left), and the ground state is partly occupied with $%
\protect\lambda_0=0.5$ (Right), respectively.}\label{jcf}
\end{figure}

We first consider the amplitude damping dynamics, it can be found in the main text that the state of system at the instant $t$ can be described by the following density matrix:
\begin{equation}
\rho_t=\left(1-e^{-\Gamma_t}\sum_{i=1}^{N-1}\lambda_i\right)\left\vert
0\right\rangle\left\langle
0\right\vert+e^{-\Gamma_t}\sum_{i=1}^{N-1}\lambda_i\left\vert
i\right\rangle\left\langle i\right\vert, \label{jc_solu}
\end{equation}
where $\lambda_i$ is the $i-$th eigenvalue of $\rho_0$.
We only consider the Markovian regime here, for simplicity, let $\Gamma_t=\gamma t$ with constant $\gamma$, and we take the case of $N=3$ as example. We numerically calculate the QSL in $\tau_\alpha$ with different initial state. The numerical result is shown as Fig. \ref{jcf}. It can be seen that the QSL in Theorem 1 is attainable when $\lambda_1=\lambda_2$, which means that the evolution trajectory can be written as the form of
\begin{equation}\label{app1}
  \rho_t=p_t\rho_0+(1-p_t)\frac{\mathbb I_N}{N},
\end{equation}
which is the geodesics in our metric space.

\begin{figure}
\includegraphics[width=0.38\linewidth]{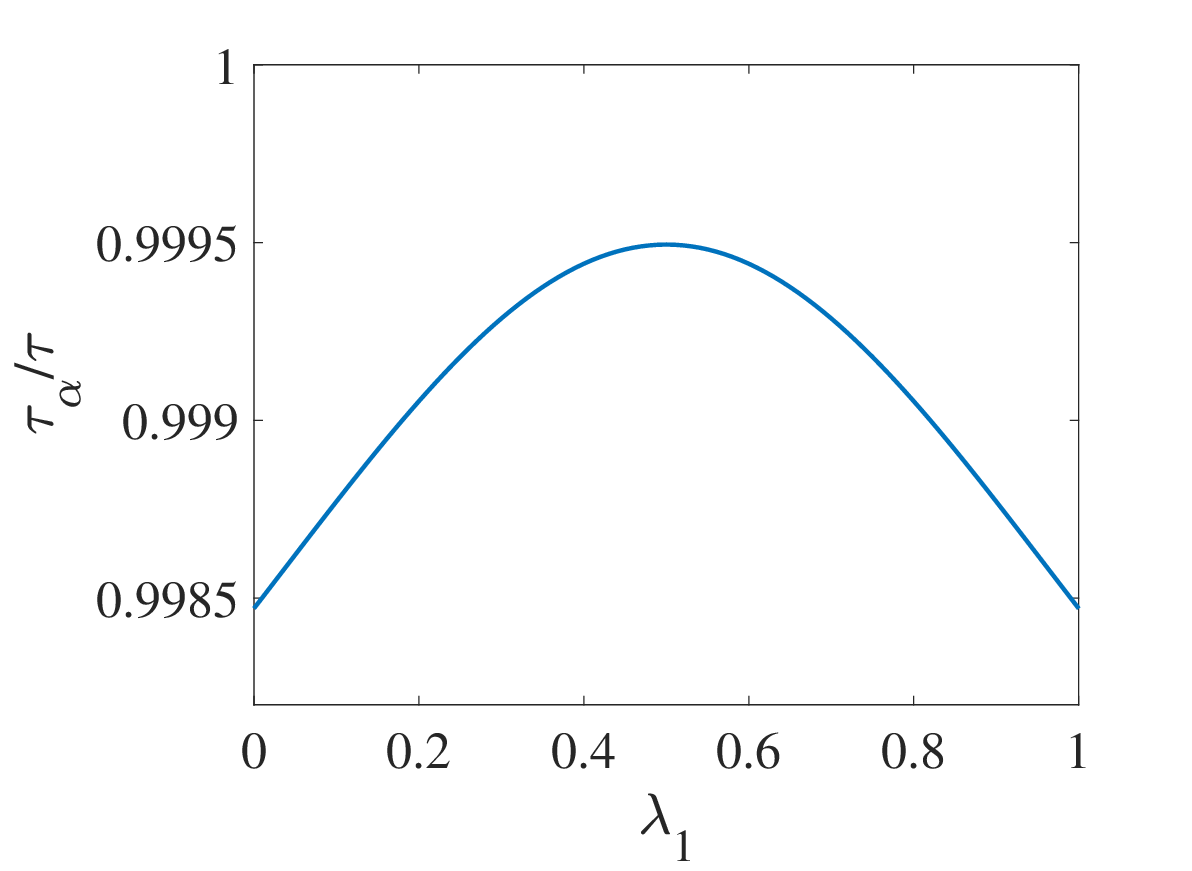}%
\includegraphics[width=0.38\linewidth]{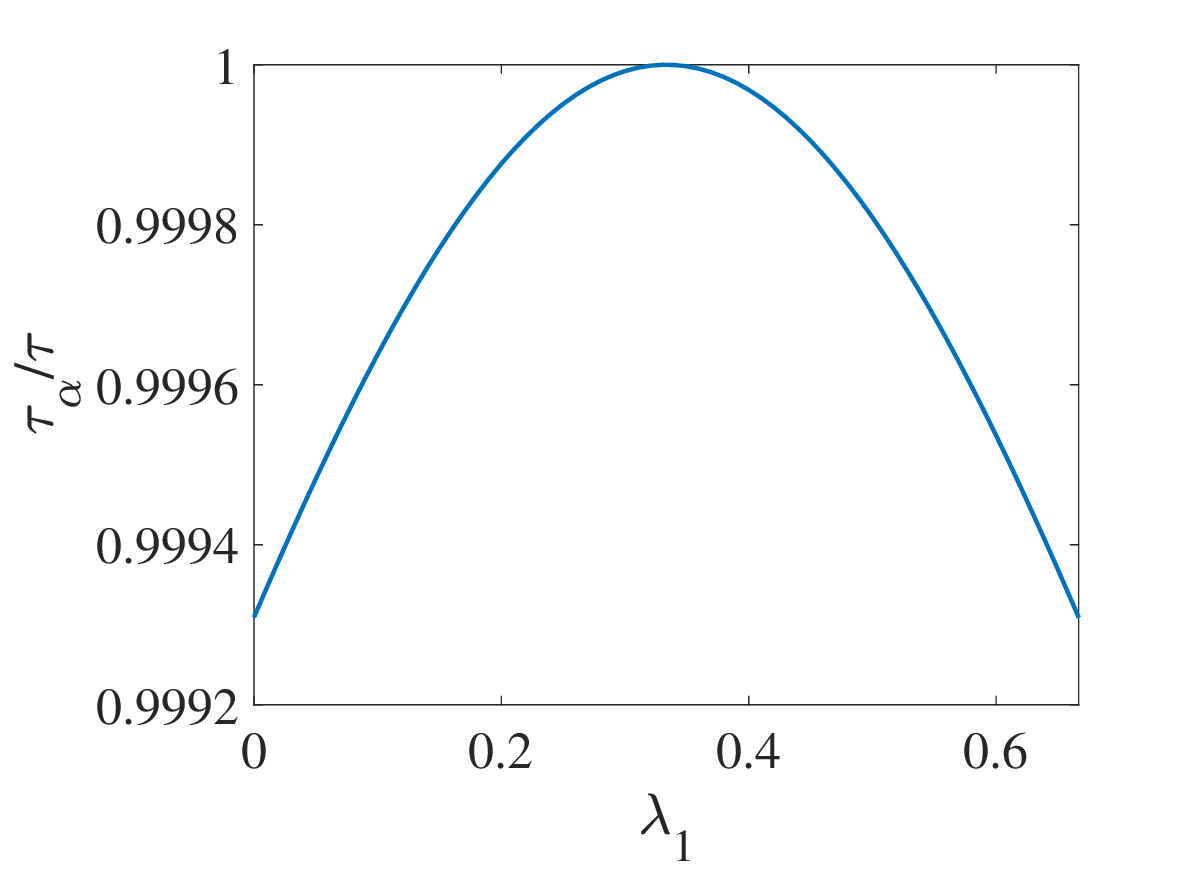}
\caption{Different initial state vs corresponding QSL. The evolution time is
$\protect\tau=1$. The system is prepared as excited state initially with $%
\protect\lambda_0=0$ (Left), and the ground state is partly occupied with $%
\protect\lambda_0=1/3$ (Right), respectively. $\lambda_{20}$ is set for ensuring the initial state is pure, and $\lambda_{ij}=0$ for the rest $i,j$.}\label{dephasf}
\end{figure}

As another application example, let's apply the QSL $\tau_\alpha$ to the dephasing dynamics: Suppose an $N-$level system driven by the internal Hamiltonian $H=\sum_{i=0}^{N-1} E_i\left\vert i\right\rangle\left\langle i\right\vert$, the system couples to the multiple reservoirs with each energy levels driven by an individual reservoir. The interaction between the reservoir and the $j,k-$th levels can be described by the Hamiltonian $H_{jk}=\sum_{\mu}\sigma_{jk}^{z}\left(g_\mu b_\mu^\dagger+h.c\right)$ with $\sigma_{jk}^z=\left\vert j\right\rangle\left\langle j\right\vert-\left\vert k\right\rangle\left\langle k\right\vert$ for $\forall j>k$, $g_\mu$ and $b_\mu$ represent the coupling strength and particular operator of the reservoir in terms of the $j,k-$th levels. Assume the system is initially prepared as $\rho_0=\sum_{i=0}^{N-1}\lambda_i\left\vert i\right\rangle\left\langle i\right\vert+\sum_{i<j}\lambda_{ij}\left\vert i\right\rangle\left\langle j\right\vert+h.c$, the system is uncorrelated with the reservoirs in the initial instant, then one can find that the reduced system undergoes the dephasing channel at the instant $t$ can be described by the following density matrix:
\begin{eqnarray}
  \rho_t &=&\sum_i\lambda_i\left\vert i\right\rangle\left\langle i\right\vert +p_t\sum_{j\neq k}\lambda_{jk}\left\vert j\right\rangle\left\langle k\right\vert
\end{eqnarray}
with $p_t=e^{-\gamma t}$, and $\gamma$ is the decay rate. It can be seen that in Fig. \ref{dephasf}, QSL is unsaturated except for $\lambda_i=1/3$ for $i=0,1,2$, which is the case that the evolution trajectory passing through the maximally mixed state, and this trajectory coincides with the geodesics. For the other cases, the unsaturated QSL reflects potentials to be speed up for the evolution.
The above two cases demonstrate the QSL $\tau_\alpha$ presented in Theorem 1 is attainable for the dissipative dynamics. However, the QSL $\tau_\alpha$ is unattainable for the unitary dynamics if $N>2$.

\begin{figure}
\includegraphics[width=0.38\linewidth]{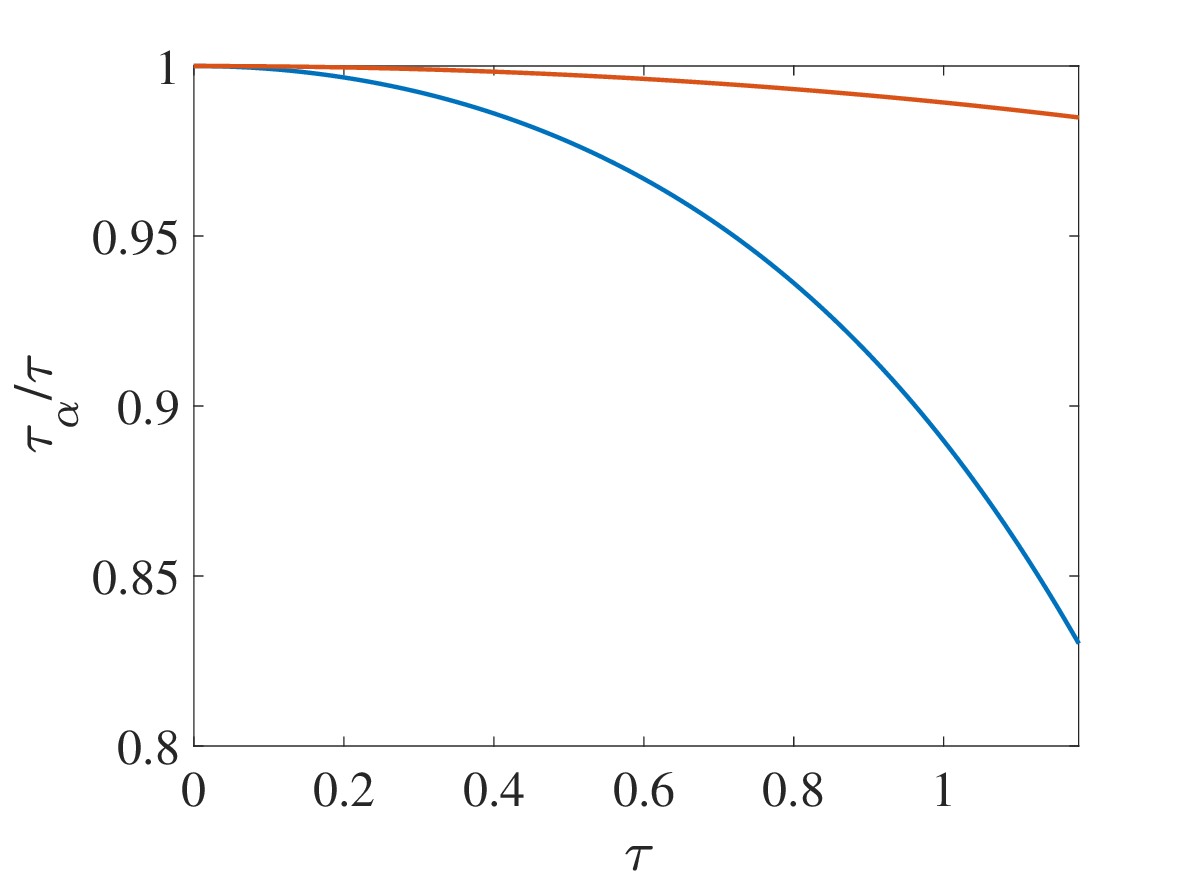}%
\caption{The QSL time $\tau_{\mathrm{QSL}}$ versus the actual unitary evolution time  $\tau$. The red line corresponds to the optimal Hamiltonian $H_T$, and the blue line means $H=H_0+H_1$.  Here $E_m=1$, $\Omega=E_m$ and $\mu=0.5$.}\label{unitaryf}
\end{figure}

To show this point, let's return to the case we presented in the main text: considering the system is initially at the ground state $\rho_0=\left\vert 0\right\rangle\left\langle 0\right\vert$, the eigenvalues of the systemic Hamiltonian is bounded by $E_m$ as $H_0=\mu E_m\left\vert 1\right\rangle\left\langle 1\right\vert +E_m\left\vert 2\right\rangle\left\langle 2\right\vert$, to drive the system to be excited, we impose external interaction to achieve the total Hamiltonian as $H=H_0+\Omega\left(\left\vert 0\right\rangle\left\langle 2\right\vert+\left\vert 2\right\rangle\left\langle 0\right\vert\right)$ and $H=H_T=U^FH_0\left(U^F\right)^\dagger$, respectively, where $H_T$ is the optimal Hamiltonian which saturates our QSL $\tau_{\mathrm{QSL}}$ in Theorem 2. We show the numerical result of $\tau_{\alpha}/\tau$ with different evolution time for the unitary evolution process in Fig. \ref{unitaryf}, it can be seen that $\tau_{\alpha}$ always deviates from the actual evolution time $\tau$, it means that $\tau_{\alpha}$ overestimate the evolution time compare with $\tau_{\mathrm{QSL}}$. However, one can still observe that $\tau_{\alpha}$ deviate less from the actual evolution time $\tau$ for the case of the optimal Hamiltonian $H_T$, it indicates that the evolution speed of closed systems driven by $H_T$ is closer to the speed limit, compare with the Hamiltonian $H=H_0+H_1$. Let's consider the unattainable properties of $\tau_\alpha$ for the unitary dynamics can be attributed to the fact as follow:  For any pairs of density matrices, the shortest unitary path connecting them is a great circle lying on a two-dimensional spherical surface, these trajectories can always be attributed to the action of a $2$ by $2$ Hamiltonian, hence for the high level systems ($N>2$), $\tau_{\alpha}$ is unattainable, this is the reason to unattainable property for many of existed QSLs as well.

\bibliography{sm}